\documentclass[preprint,12pt,authoryear]{elsarticle}

\usepackage{lineno,hyperref}
\modulolinenumbers[5]

\usepackage{float}
\usepackage{amssymb}
\usepackage{color, colortbl}
\usepackage{soul,xcolor}

\usepackage{makecell}
\usepackage{subfigure}

\usepackage{import}
\usepackage[utf8]{inputenc}
\usepackage[english]{babel}

\journal{Nuclear Physics A}

\begin{document}

\begin{frontmatter}

\title{Optimization study of the electrode design of a 5 mm thick orthogonal-strip CdZnTe detector system}
\author[sabanci]{Ali Murteza Alt\i ng\"un\corref{mycorrespondingauthor}}
\cortext[mycorrespondingauthor]{Corresponding author. Ali Murteza Alt\i ng\"un}
\ead{aaltingun@sabanciuniv.edu}

\author[sabanci]{Emrah Kalemci}
\ead{ekalemci@sabanciuniv.edu}

\address[sabanci]{Faculty of Engineering and Natural Sciences, Sabanc{\i} University, Orhanl{\i}, Tuzla, Istanbul, 34956, Turkey}

\begin{abstract}
      The geometry of electrodes is one of the most important factors in determining the performance of orthogonal-strip detectors. The aim of this work is to study the performance of a 5 mm thick cross-strip CdZnTe detector with different electrode widths. Our study consists of two main parts, simulations and experiments. We utilized four different anode sizes ranging from 0.1 mm to 0.6 mm. The anodes were interspersed with steering electrodes with varying sizes from 0.3 mm to 0.85 mm. The maximum gap size between the anodes and steering electrode strips was set to 0.3 mm, while the minimum gap size was 0.125 mm. The performance of the detector was investigated in terms of the steering electrode bias voltage, the energy resolution, and the charge sharing effect. For simulations, we developed a C++ based simulation program for charge transport inside the CdZnTe detector and charge collection at the electrodes. For photon interactions we used GEANT4 toolkit and for electric field and weighting potential simulations we used COMSOL software. The results demonstrated that -50 V is the optimal steering electrode bias for our detector when -500 V was applied to the cathodes and that the energy resolution performance drops with increasing steering electrode width. Also, the charge sharing effect becomes more dominant for larger steering electrode sizes. The experimental result are further compared with the simulations. The results are in a good agreement and the comparison validates our simulation model. Although, our simulation framework has need of better estimation for the intrinsic noise of CdZnTe. These results suggest that an optimization study between electrode widths and steering electrode bias is required to obtain the best performance in orthogonal-strip CdZnTe detectors.
\end{abstract}

\end{frontmatter}

\section{Introduction}
\label{sec:intro}
In recent years, important advantages of Cadmium Zinc Telluride (CdZnTe) based semiconductor detectors have attracted the attention of many researchers in the fields of medical imaging, astronomy, and homeland security. Due to its high stopping power, high internal resistivity, and large band gap providing a good energy resolution at room temperature \cite{Kalemci02}, CdZnTe outperforms most other types of hard X-ray detectors. Many studies with different geometries have shown that CdZnTe detectors possess very good energy \cite{Zhang_2005}, \cite{kuvvetli_2014}, \cite{Abbaszadeh_2016}, spatial  \cite{Abbaszadeh_2016}, \cite{Zheng_2016} and time resolution characteristics \cite{Abbaszadeh_2016}, \cite{Meng_2005}, \cite{Zhao_2012}. Pixellated and cross-strip electrodes are the two prominent CdZnTe detector designs in use. In the pixel design, anodes are composed of small electrodes (pixels), while the cathode is a whole planar electrode. This configuration provides quite good spatial resolution, as well as energy resolution due to the small pixel effect. However, the readout electronics cost increases with the number of pixels since every pixel requires a single readout channel. In the cross-strip detector design, on the other hand, anodes and cathodes are perpendicular to each other, imitating NxN pixels with 2xN electronic channels. This promises to be especially advantageous for applications that require less power and weight, such as for small satellites in space \cite{Kalemci13}, \cite{kuvvetli_2014}. The reduced complexity of the readout electronics and low cost make cross-strip configurations advantageous also for Compton cameras \cite{Kormoll_2011}, \cite{Gonzalez_2014}, \cite{Owe_2019}, \cite{Abbene_2020} and positron emission tomography (PET) applications that utilize large number of detectors \cite{Matteson_2008}, \cite{Peng2010}, \cite{Gu_2014}, \cite{Abbaszadeh_2017}. While the cross-strip design decreases the number of required readout channels drastically, due to higher capacitance, the energy resolution is poorer compared to the pixellated designs.
 
A photon interaction in CdZnTe volume creates $ N = \frac{E_{\gamma}}{\varepsilon}$ electron-hole pairs on average (analogous to electron-ion pairs in gaseous detectors), where $E_{\gamma}$ is the incident photon energy and $\varepsilon$ = 4.6 eV is the pair creation energy, the energy required to create an electron-hole pair in the crystal. 
As bunch of electrons (holes) produced by a single photon, which is called an electron (a hole) cloud, drift toward the anodes (cathodes) by the applied electric field, its size will expand due to carrier diffusion and charge repulsion. Because of the expansion in the cloud sizes, the charge carriers may induce charges on several neighboring electrodes \cite{Kalemci02}.This phenomenon is called charge sharing effect which worsens the spectral performance of the detectors. However, there are methods available to mitigate the charge sharing effect to recover good spectral performance \cite{BUGBY2019}, \cite{Abbene2018}.

Despite its promising properties, CdZnTe detectors suffer from hole collection inefficiency. Due to the low mobility of holes and the defects in CdZnTe crystals, holes are trapped. This results in a low energy tail in the pulse height spectrum. By utilizing the small pixel effect \citep{Barret}, \citep{Luke_1996}, it is possible to make the detectors sensitive to only electron motion. The small size of a pixel or strip ensures the weighting potential of the anodes to have near-zero values in the vicinity of the cathode. Therefore the motion of holes will induce negligible charge on the anodes.

Some electrons might drift to regions between the anodes, which causes incomplete electron collection. To overcome this problem, anodes can be interspersed with steering electrodes (SE), which is used in CdZnTe based detectors succesfully for almost two decades. By applying a potential between the anodes and the SE, electrons are steered toward the anode strips,  resulting in an improvement in the electron collection efficiency and the spectral performance \cite{Kuvvetli_2005}, \cite{Abbene_2007}, \cite{Abbene_2020}.  Also, steering electrodes create an electrostatic shielding so that the hole collection inefficiency does not affect the spectral performance of the detectors. The effect of different potentials between the anodes and the SE in strip detectors has been reported in \cite{Gu_2014}, \cite{Abbaszadeh_2017}.

In the cross-strip design, the sizes of the strips and the gaps between them affect the detector performance significantly \cite{Gu_2014}. We performed experiments with a CdZnTe detector that has anode and SE strips with varying sizes in addition to uniform size cathode strips aligned orthogonal to the anodes. The results of the experimental work were then compared to simulations with the motivation of offering an optimization study for the cross-strip CdZnTe detectors that can be utilized for applications that require low cost, power, and complexity. Some of the earlier results of the optimization study presented here have already been used in the design of the XRD cross-strip CdZnTe detector on BeEagleSat \cite{Kalemci13}.

\section{CdZnTe crystal with orthogonal strip configuration and associated readout electronics}
\label{sec:czt_readout_electronics}
\begin{figure}
\begin{center}
\includegraphics[width=1.0\textwidth]{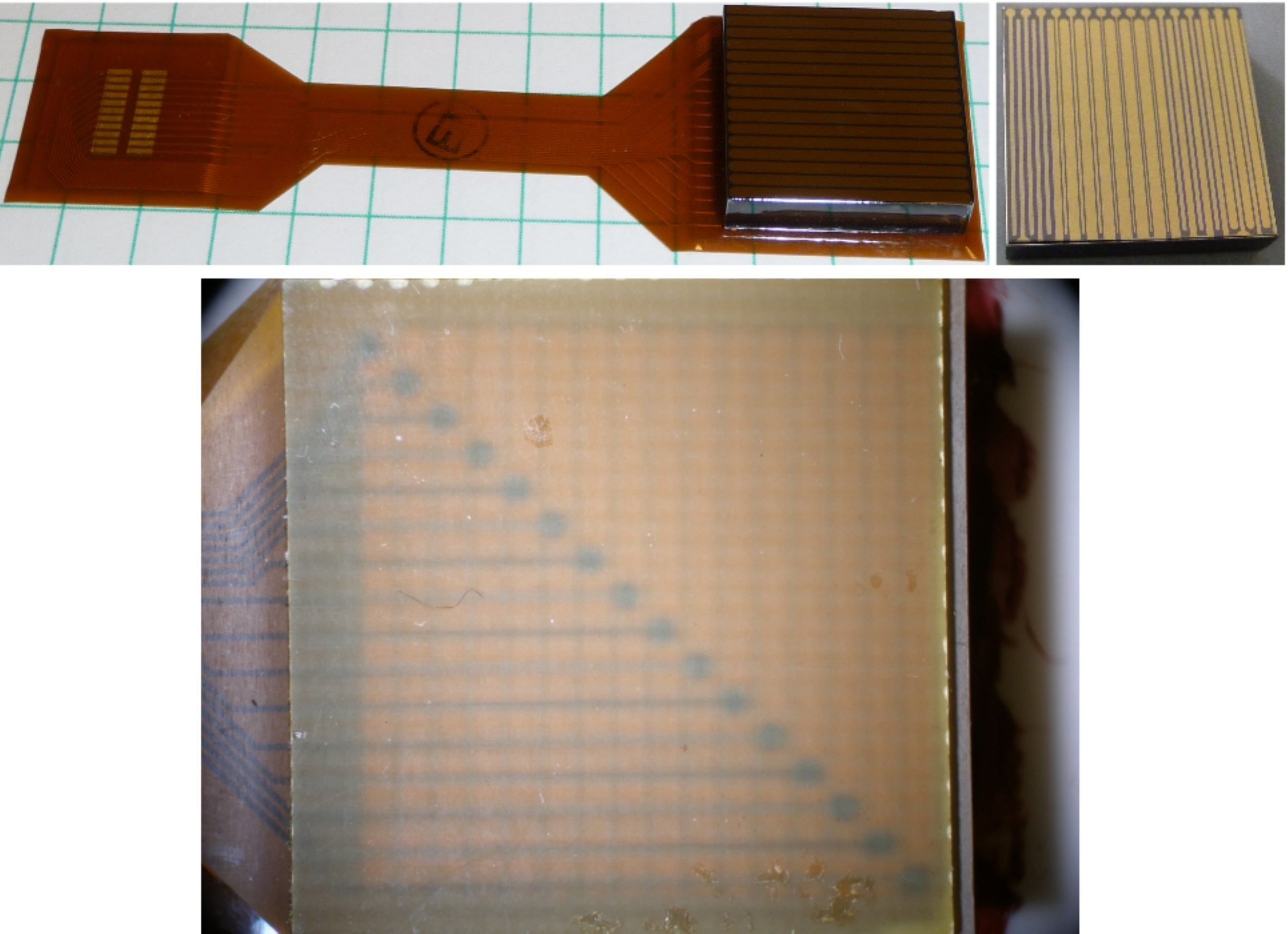}
\caption{Left top: Picture of the CdZnTe crystal, anodes attached to the flex+rigid PCB, cathodes are still visible. Right top: Anode side of the detector before flex+rigid PCBs attached. Bottom: Cathode side of the detector attached to the flex+rigid PCB.}
\label{fig:cztflex}
\end{center}
\end{figure}

We used a detector grade \cite{Duff2008} CdZnTe crystal produced by REDLEN Technologies Inc. in Canada. The dimensions of the crystal are 19.54 x 19.54 x 5 mm$^3$ (see Fig.~\ref{fig:cztflex})\footnote{The detector was purchased in 2011 and the data was taken shortly thereafter in 2012.}. Polishing, passivation and contact deposition using gold were done at Creative Electron, Inc. in San Marcos, California, USA. The crystal was then sandwiched between two flex+rigid printed circuit boards (PCBs). The signals from the anodes, SE and cathodes go through the connectors at the end of the flex PCB to a filter board for AC coupling to the RENA 3 ASIC (Application Specific Integrated Circuit) for readout. The details of the readout circuitry are given in \S\ref{subsec:rena}.

\subsection{Strip geometry description}

\begin{figure}
\begin{center}
\includegraphics[width=1.0\textwidth]{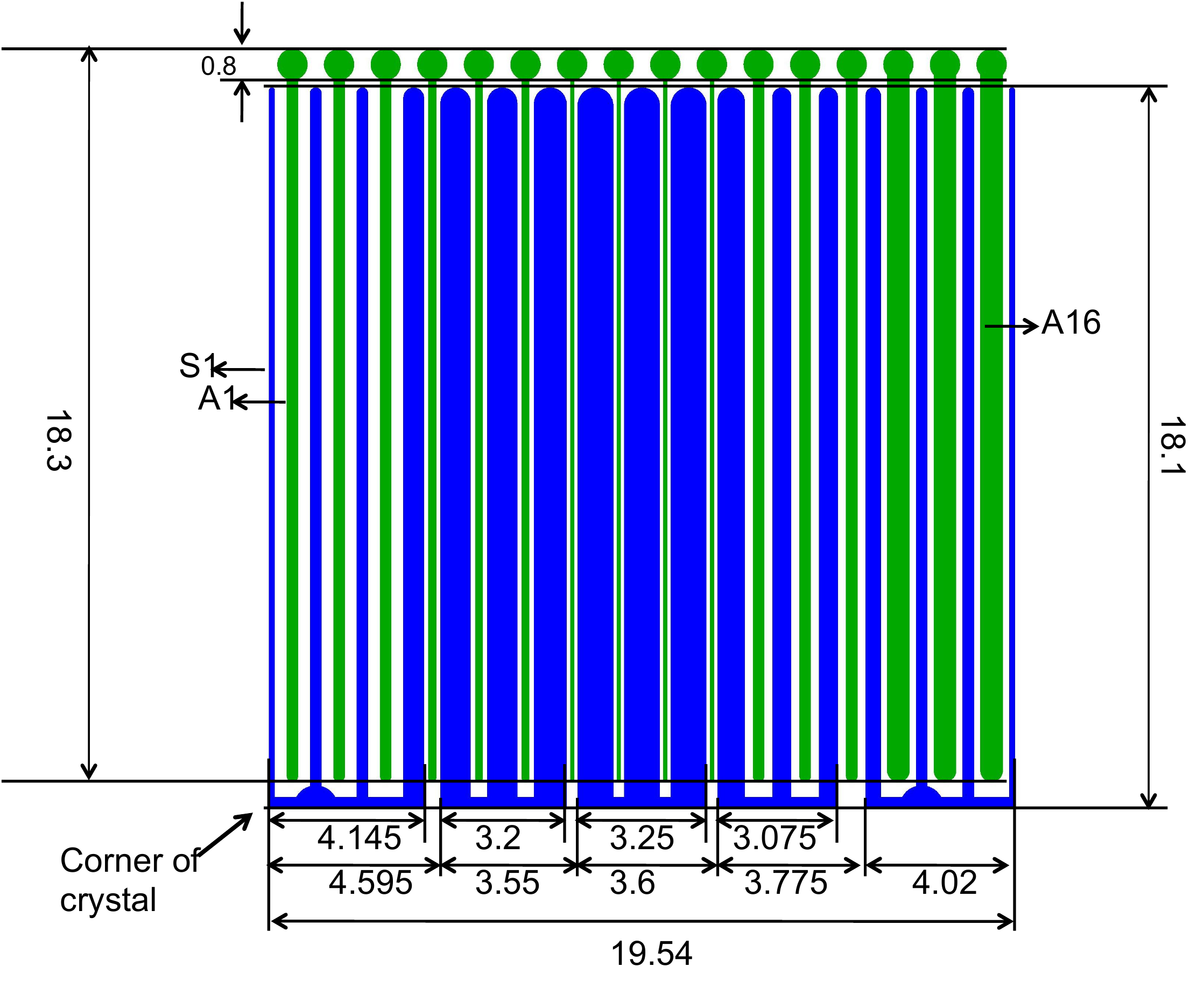}
\caption{Drawing of the anode/SE side. Green strips represent the anodes, while the blue ones are for the SE strips. Pitch is constant at 1.2 mm. All measurements are in mm.}
\label{fig:anodes}
\end{center}
\end{figure}

To study the effect of strip geometry on detector performance, we designed a multi-anode / multi-steering electrode geometry keeping the pitch constant at 1.2 mm (see Fig.~\ref{fig:anodes}). Overall, there are 16 anodes and 5 SEs in between the anodes. The SEs divide the surface to 5 sections with the same anode and SE widths  so that the anode/SE combination at the middle of each section has a well defined electric field and weighting potential representative of that section. The anode/SE/gap widths at the middle of each section are given in Table~\ref{table:widths}. For easier connection to the PCBs, we employed circular pads on one side of the anodes. We used 35 channels of the RENA ASIC. Of these, 16 were for anodes, 16 were for cathodes and the remaining 3 were for SE at a time. To complete the SE dataset, we had two runs of data collection with different SE configurations.

We used 16 orthogonal cathode strips which are uniformly laid out, again with a pitch of 1.2 mm. Since the cathodes should be made larger to collect enough charge at larger depths \citep{Kalemci02}, we used 1 mm wide electrodes with 0.2 mm gaps in between. Both anode side and cathode side electrodes are 0.27 mm away from the sides. There is no guard ring. 
     
\begin{table}
	\caption{Geometry of anodes and steering electrodes}
	\label{table:widths}
	\begin{tabular}{ccccc}
Group$^{1}$ & Anode & SE & Gap & Assoc. Anodes$^{2}$ \\ 
 & mm & mm & mm & \\ \hline
1 & 0.3 & 0.3 & 0.3 & A1, A2, A3 \\ 
2 & 0.2 & 0.75 & 0.125 & A4, A5, A6 \\ 
3 & 0.1 & 0.85 & 0.125 & A7, A8, A9, A10 \\ 
4 & 0.3 & 0.5 & 0.2 & A11, A12, A13 \\ 
5 & 0.6 & 0.3 & 0.15 & A14, A15, A16 \\  \hline
\multicolumn{5}{l}{$^{1}$ Steering electrodes are named after their groups such as SE1, SE2, etc.}
\\
\multicolumn{5}{l}{$^{2}$ The anodes are named as A1, A2, A3, etc.}
\end{tabular}
\end{table}
     
\subsection{Readout circuitry}
\label{subsec:rena}
     
\begin{figure}
\begin{center}
\includegraphics[width=1.0\textwidth]{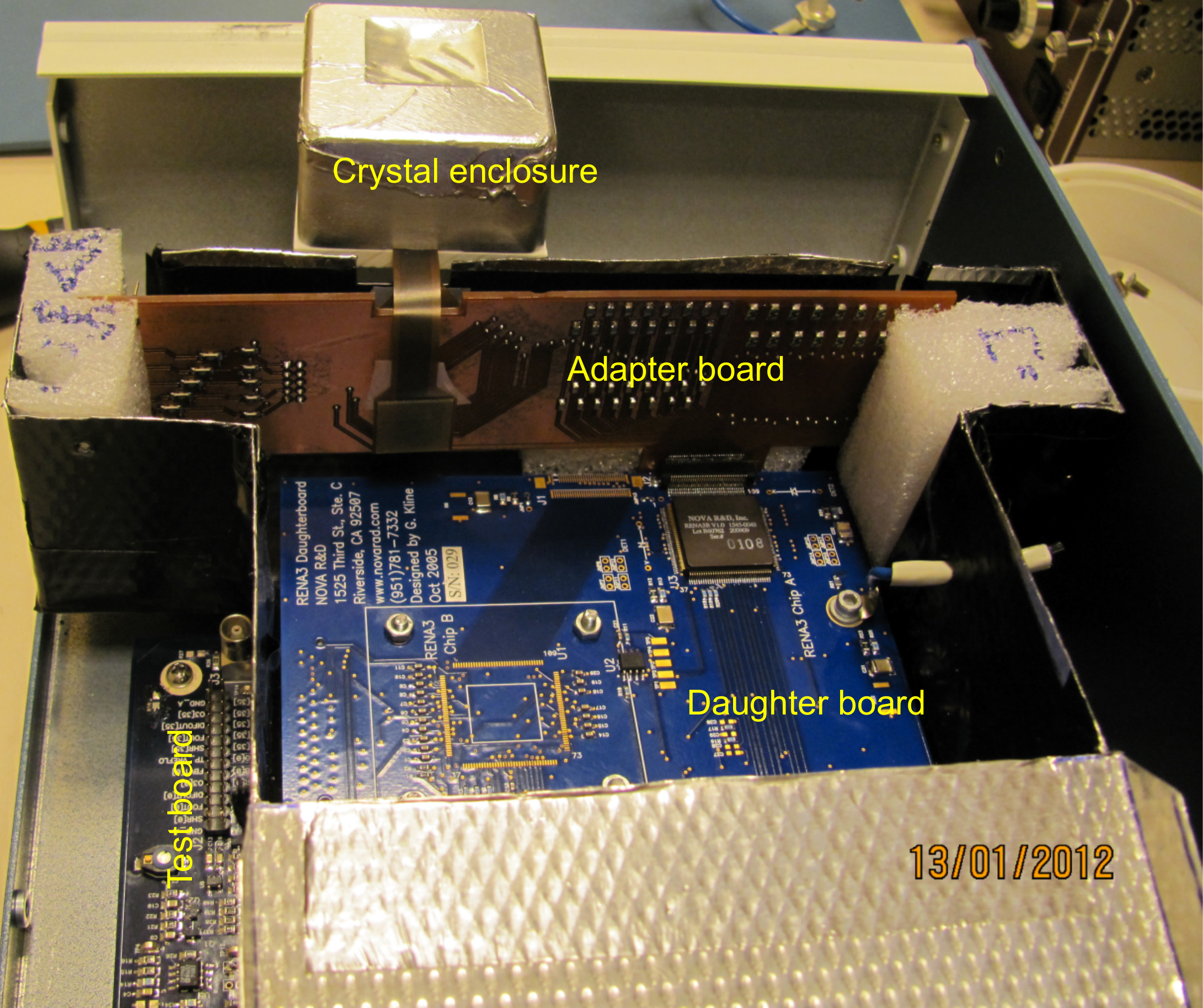}
\caption{Picture of the experimental system. During data acquisition the system is completely enclosed in metal to minimize noise.}
\label{fig:systempic}
\end{center}
\end{figure}

We utilized a single RENA 3b, which is a low-noise, 36-channel, self-trigger, self-resetting charge sensitive amplifier/shaper ASIC developed by Nova R\&D (now part of Kromek Group) \cite{Tumer08}. It is capable of reading both anode and cathode signals. The RENA is a part of a daughter-board that attaches onto a main evaluation board developed by Nova R\&D (see Fig.~\ref{fig:systempic}). The ASIC is highly configurable, and based on our anode and cathode geometry and crystal thickness we operated RENA 3b with these settings: 1.9 $\mu$s shaping time, 15 fF feedback capacitance (the low energy range, 9fC/256 keV), anode feedback resistance of 1200 $\Omega$, cathode and SE feedback resistance of 200 $\Omega$, and a FET size of 450 $\mu$m for anodes and 1000 $\mu$m for cathodes and SE. We also operated a pulser to measure the electronic noise. The pulser amplitude is chosen to be close to the prominent lines of $^{57}$Co.

The AC coupling and the powering of the crystal were performed with the help of an adapter circuit board which can be attached to the daughter-board. All electrodes were coupled to RENA with 1 nF coupling capacitors. We utilized AC coupling because the measured leakage current per strip (1 - 6 nA) is close to or higher than the tolerance limits of RENA ASIC \cite{Tumer08}.  The power was provided externally with an ORTEC 556 high voltage power supply through the adapter board. Generally we applied -500 V to the cathodes and -50 V to the SE through 100 $M\Omega$ resistors. The anodes were grounded with 100 $M\Omega$ resistors as well. The SE voltage was produced by a simple voltage divider circuitry. The RENA ASIC is configured to trigger on negative signal for the anodes and the SEs, and trigger on positive signal for the cathodes. A picture of the experimental system is shown in Figure \ref{fig:systempic}. For a better noise performance, the crystal was enclosed in an aluminium box. The aluminum enclosure has a window for the source irradiation. This window was closed  with a thin aluminum tape. The daughter board was also enclosed in a metal cage to minimize electronic noise caused by digital circuitry in the evaluation board.
\section{Experiments}     
\label{sec:experiments}
  We have conducted measurements using $^{57}$Co (122.1 keV, 136.5 keV) and $^{241}$Am (13.9 keV, 17.8 keV, 59.6 keV). For each source we have around 8 x $10^{5}$ registered events. A pulser was employed to determine the electronic noise. The calibration to obtain photon energies from the RENA ADC values was conducted for all electrodes using 59.5 keV and 122.1 keV lines. We only used the events that we denote as singles; no charge sharing among the anodes and the steering electrodes, events ending up on a single anode and an accompanying signal on cathode(s). We calibrated the SE in the same way using events collected by a single SE with an accompanying signal on cathode(s). Since the cathode strips are wide and sensitive to larger depths, it is sometimes required to add the cathode signals to obtain total charge collection. Depth of interaction in CdZnTe can be deduced by using the ratio of the cathode to anode signals \cite{DOI_W_Li_1999}. Signals produced by the interactions occurred in the proximity of the cathode side will be affected less by the hole collection inefficiency, since the anode weighting potential values are nearly zero in the regions close to the cathodes. 
  For calibration, we analyzed anode (cathode) signal ADC channel vs. cathode to anode ADC channel ratio plots for each anode (cathode) strip, and used events within a $\pm$10 \% ratio region around the peak that corresponds to full charge collection. After calibration, the cathode to anode ratio for each single event was calculated by dividing the calibrated anode signal to the sum of calibrated cathode signals.
 
  The minimum detectable energy of the system depends on the internal properties of the CdZnTe crystal, the electrode configuration and goemetry, as well as the noise properties of the readout circuitry. In RENA 3b, each channel has a trigger comparator circuit. A threshold value is set for the comparators according to the noise level of the entire system. We checked each channel and determined trigger thresholds such that the pedestal due to electronic noise produced a background rate much less than the actual count rate. Threshold values of 20 keV, 35 keV and 40 keV were set for the anode, cathode and SE channels respectively. During testing, cathodes numbered 1-8 disconnected due to a problem with the cold-solder. Similarly, A9 produced no signal due to a break in the connection. For the calibration, and the energy resolution measurements per strip, we only used the singles inducing signals only on one of the cathodes numbered 9-16 to minimize any possible adverse affect caused by unconnected cathodes. Whether the disconnected electrodes affect the results will be discussed in the section  \S~\ref{sec:sim_charge_sharing}
\section{Simulations}     
\label{sec:simulations}
\begin{figure}
\begin{center}
\includegraphics[width=0.8\textwidth]{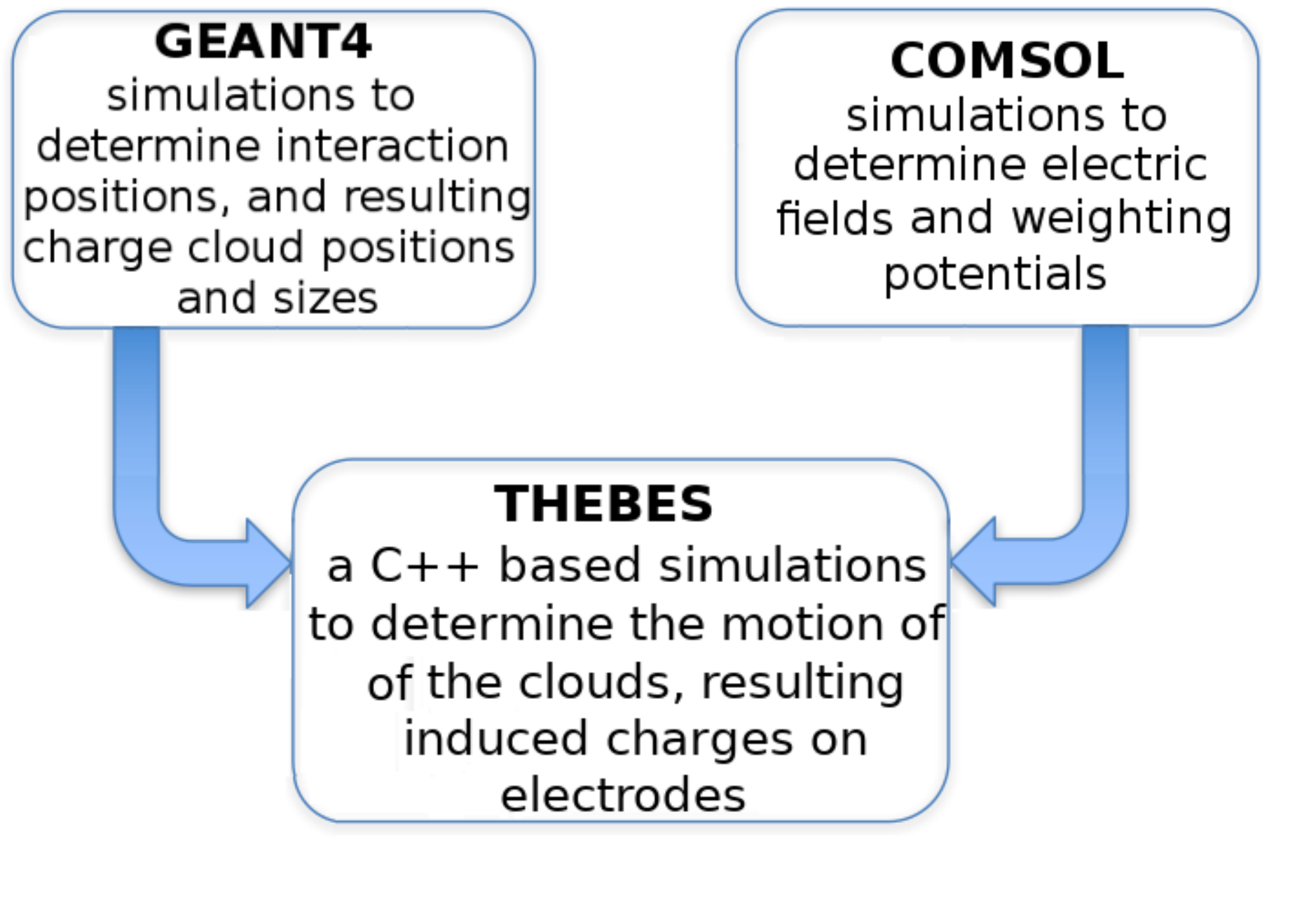}
\caption{Simulation method.}
\label{fig:simethod}
\end{center}
\end{figure}

 To analyze the detection characteristics of different widths of electrodes, we conducted simulations of charge generation and collection in our CdZnTe detector system and compared our experimental results with the predictions of the simulations. The simulations include 3 main steps as shown in Figure~\ref{fig:simethod}.
 
 The general method of simulating charge induction and collection inside the crystal is described in \cite{Kalemci99} where we track charges inside the detector by moving them along the electric field lines and find induced charges using the weighting potentials method \citep{HE_2001}. 

\begin{figure}
\begin{center}
\includegraphics[width=0.8\textwidth]{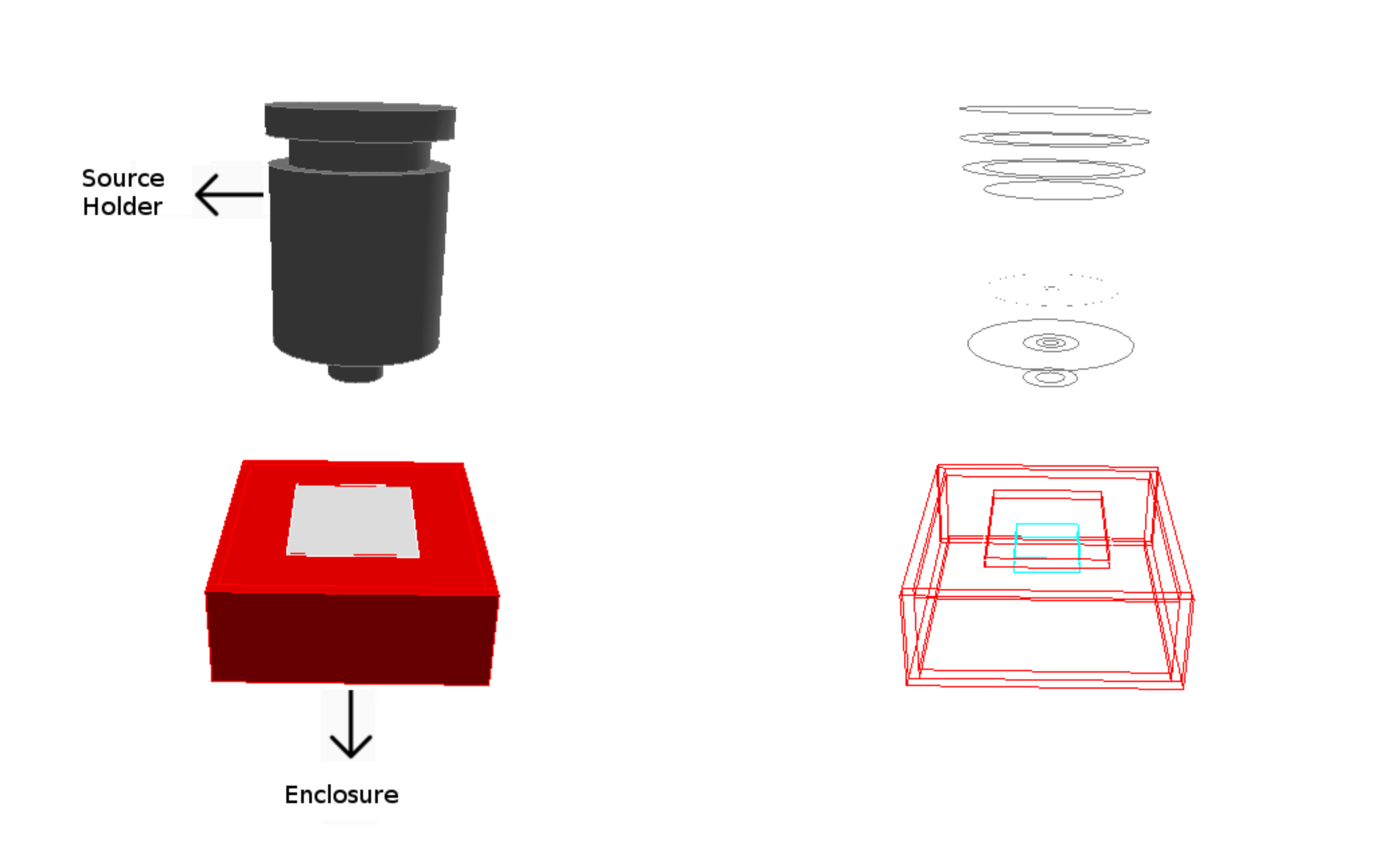}
\caption{Visualization of the GEANT4 Simulation. Crystal (magenta color) is visible in the aluminum box. A 20 micron thick aluminum plate is placed in front of the enclosure window (light gray color) }
\label{fig:g4sim}
\end{center}
\end{figure}

 
 By using GEANT4 Toolkit \cite{Agostinelli_G4_2003}, we have determined the interaction positions of photons in the crystal in 3D as well as the energy of initial charge clouds created by the incoming photons. We used emstandard$\_$opt4 physics constructor to simulate the interactions of the photon and matter. The radioactive calibration source is in an aluminum holder surrounded by lead cylinders and photons are permitted from the tip with an opening diameter of 4 mm. The crystal is embedded in an aluminum box having a window to mimic the experimental setup (Figure \ref{fig:systempic}). A snapshot from the Geant4 simulation toolkit is given in Figure \ref{fig:g4sim}. 
 
\begin{figure}
\begin{center}
\includegraphics[width=1.0\textwidth]{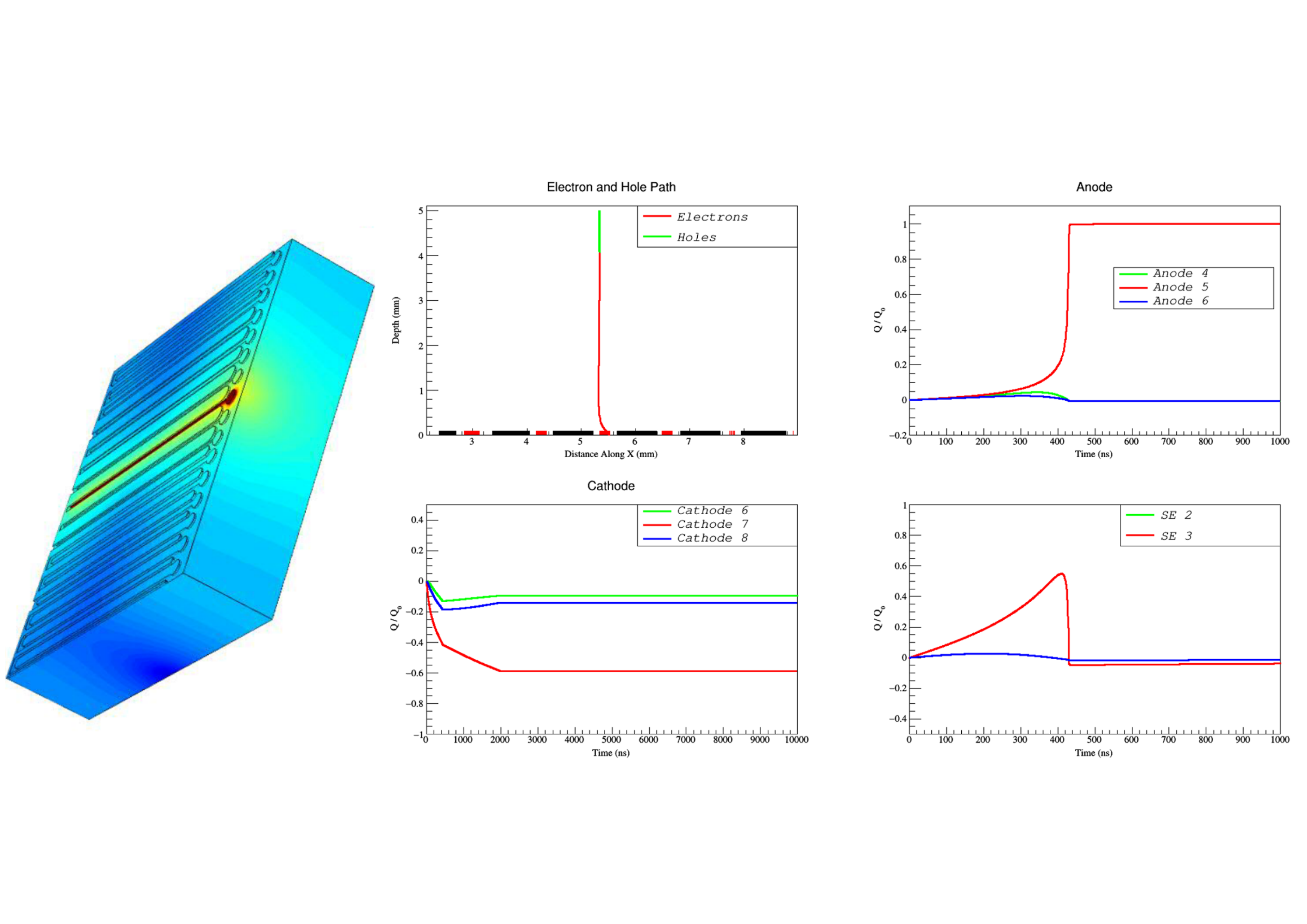}
\caption{Left: COMSOL 3D simulation of the weighting potential distribution of anode numbered 8. Right: A representative example of charge cloud tracking and charge generation for an  interaction close to cathodes, shared by 2 cathodes and ended on a single anode.}
\label{fig:4_panel_event_3_normalized_COMSOL}
\end{center}
\end{figure}

 We used COMSOL Multiphysics\textsuperscript{\textregistered} software to model electric fields and weighting potential distributions in the crystal volume by utilizing the AC/DC module. 
 To calculate the weighting potential of an electrode, its potential were set to 1 V and the rest were grounded. The COMSOL software allows the users to set different mesh sizes in different part of the geometry. Since both the weighting potential and the electric field distributions get complicated near the electrodes, extremely fine tetrahedral meshes (maximum element size is set as 0.1 while the minimum elements size is 0.01 in custom settings) were employed in the vicinity of the electrode strips to get more precise results. In the other part of the volumes, coarser mesh sizes were used. The output files were obtained by utilizing 0.01 mm step sizes for X, Y and Z coordinates. An example for the weighting potential distribution of one of the anodes is shown in Figure~\ref{fig:4_panel_event_3_normalized_COMSOL}.  

Transportation of charge clouds and charge induction on the electrodes are calculated using a C++ code (called THEBES - \textbf{T}ransporter of \textbf{H}oles and \textbf{E}lectrons \textbf{B}y \textbf{E}lectric field in \textbf{S}emiconductor Detectors) we developed. THEBES utilizes the results from the GEANT4 and COMSOL software. THEBES allows users to choose mobility and life-time for the electrons and the holes, geometry of the detector and the electrodes as input. The crystal volume is divided into cubic grids with 0.01 mm sides. Each grid element has electric field and weighting potential values from COMSOL. Geant4 simulation provides the positions of each charge cloud. A charge cloud moves one grid perpendicular to the electrodes and the time it takes for this one grid-motion of the cloud is calculated by using the method reported in \citep{Kalemci99}. The lateral propagation of the cloud is computed in this time. This process continues until the charge clouds reach the electrodes or are completely trapped. 
For each grid point, the induced charges on all electrodes are calculated and saved. In the simulation, the electron cloud expansion is also considered. As the electron cloud moves toward anode strips, the size of the cloud expands due to charge diffusion and electrostatic repulsion. We assumed that the charge clouds are spherically distributed. The diffusion is a 3-d random walk process. In each step, each charge element is located in the x,y and z positions that are randomly selected from a Gaussian distribution.

\begin{figure}[!htb]
\begin{center}
\includegraphics[width=1.0\textwidth]{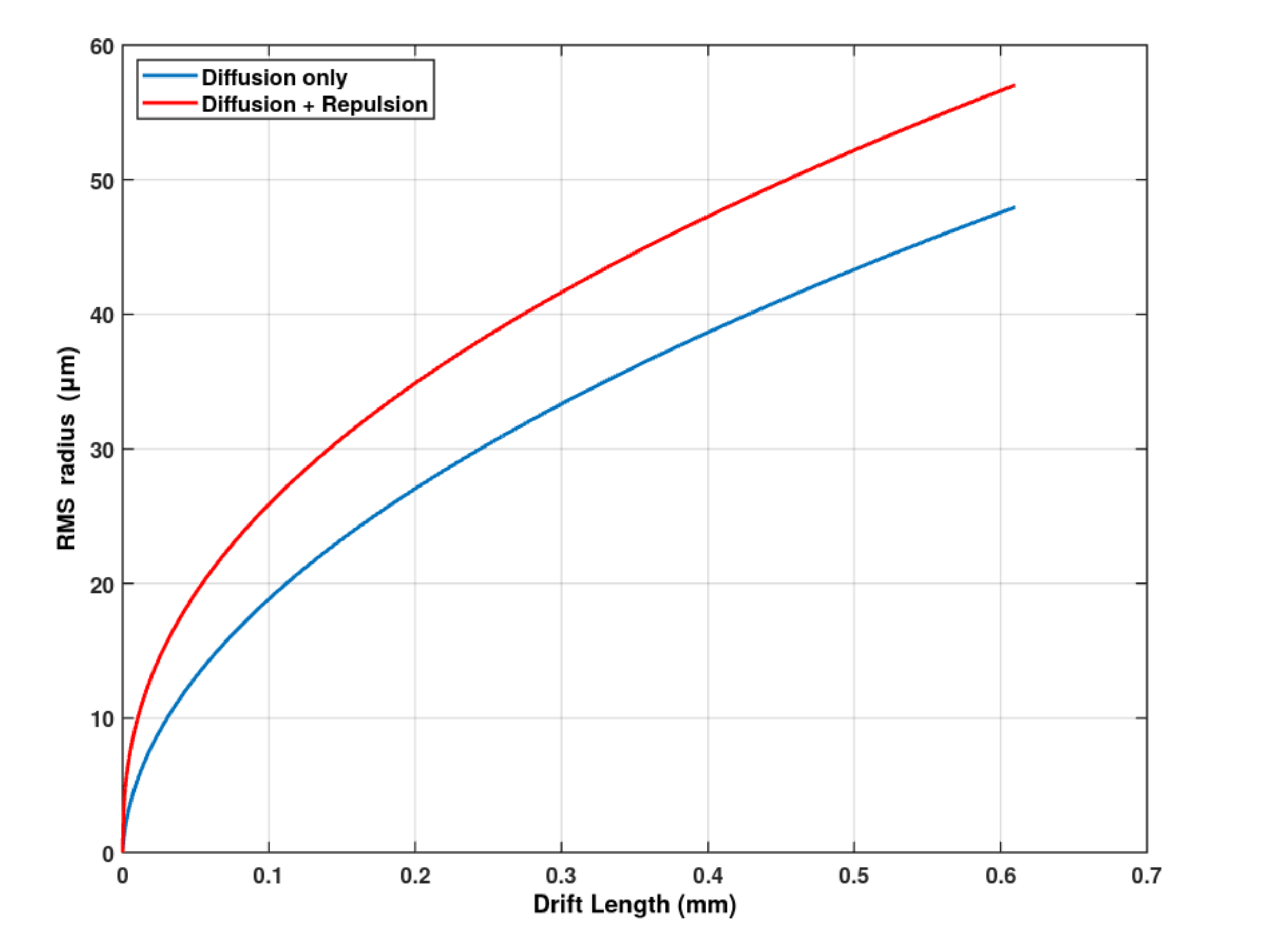}
\caption{RMS radius of electron cloud at a 122 keV energy deposition with respect to drift length. 
\label{fig:charge_cloud_expansion}}
\end{center}
\end{figure}

The equation determining the size of the charge distribution over time due to the electrostatic repulsion is given by \cite{Benoit09}

\begin{equation}
 \frac{d\sigma(t)\textsuperscript{2}}{dt} = 2D + \frac{\mu_{e} Ne}{12\pi^{(3/2)}\epsilon_{0}\epsilon_{r}\sigma(t)}
 \label{eq:diffusion_eq}
\end{equation}

where $\mu$\textsubscript{e} is the mobility of electrons, $N$ is the number of electron-hole pairs created by the incoming photon, $e$ is the electric charge, $T$ is the temperature (in Kelvin), $\epsilon_{0}$ is the vacuum permittivity and $\epsilon_{r}$ is the relative permittivity of CZT crystal. $D$ is the  diffusion constant given by

\begin{equation}
 D = \frac{\mu\textsubscript{e} k_{B} T}{e}
 \label{eq:diffusion_coeff}
\end{equation}

where $k_{B}$ is the Boltzmann constant. Equation (\ref{eq:diffusion_eq}) is solved by using the Runge-Kutta method. Figure \ref{fig:charge_cloud_expansion} shows the change in the radius of an electron cloud at a 122 keV energy deposition for different drift lengths according to Equation \ref{eq:diffusion_eq}. 

Finally, a Gaussian random noise proportional to the induced charge has been added to immitate the effect of electronic noise and other noise contributions in the crystal. As done in the experiments, threshold values of 20 keV, 35 keV and 40 keV were set for the anode, cathode and SE channels, respectively.

A representative set of parameters used in the simulation are summarized in Table~\ref{table:sim_params}. 

\begin{table}[!htb]
\centering
 \caption{Parameters used in the simulation } 
  \begin{tabular}{c c } 
 Parameters & Values  \\ [0.5ex] 
 \hline
 Electron mobility (mm\textsuperscript{2}/V.s) & 1.1 x $10^{5}$  \\ 
 Electron trapping time (s) & 5.0 x $10^{-6}$ \\
 Hole mobility (mm\textsuperscript{2}/V.s) & 10.5 x $10^{3}$ \\
 Hole trapping time (s)  & 3.0 x $10^{-6}$ \\
 Electronic noise level for anodes & 15\% \\
 Electronic noise level for cathodes and SE & 18\% \\ [1ex] 

 \hline
 \end{tabular}
 \label{table:sim_params}
\end{table}
In THEBES, the response of the readout electronics to the incoming signal is also considered. The induced charges created in the detector is converted into a current pulse. Since the amplitude of the pulses are too small, they are amplified by a charge sensitive preamplifier (CSA).
The impulse response of the CSA is given as:

\begin{equation}
             V_{CSA,out}(t) =\frac{Q_{in}}{C_{f}} \frac{\tau_{f}}{\tau_{r} - \tau_{f}} \left( e^{-\frac{t}{\tau_{r}}} - e^{-\frac{t}{\tau_{f}}}\right)
             \label{function_csa}
\end{equation}

where $\tau_{r}$ and $\tau_{f}$ rise time and fall time constants for the amplifier. $Q_{in}$ is the induced charge created by a single quantum and $C_{f}$ is the feedback capacitor for the preamplifier. Afterwards, the signal output from CSA is carried into the shaper amplifier with an output response function as:

\begin{equation}
              V_{Shaping,out}(t) =\frac{1}{n!} G^{n} n^{n} \left( \frac{t}{\tau}\right)e^{-\frac{t}{\tau}}
              \label{function_shaper}
\end{equation}
where $\tau$ is the peaking time, 1.9 $\mu$s for our case, $G$ is the shaper gain which is set to 5, and the $n$ value indicates how many RC circuits are employed after CR circuit in the amplifier architecture. The amplitude of the shaping amplifier output is used for the reconstruction of the incoming particle energy as RENA utilizes a peak \&\ hold circuitry whose output is read by an analog to digital converter . The results relevant to the objectives of this work are discussed in the next section. 

\section{Results}
\label{sec:results}

We analyzed the experimental results in terms of energy resolution, bias  voltage performance and the charge sharing effect for each set of electrode geometries. 
We applied a cathode to anode ratio based event selection and used events within the ratio range of 0.2--1.1 for each strip. This selection is determined by investigating the anode signal vs. cathode to anode ratio plots. This selection resulted in discarding 10\% to 20\% of all recorded events for each anode strip. 

\subsection{Energy resolution measurements}
\label{sec:en_res_results}
\begin{figure}[!htb]
\begin{center}
\includegraphics[width=1.0\textwidth]{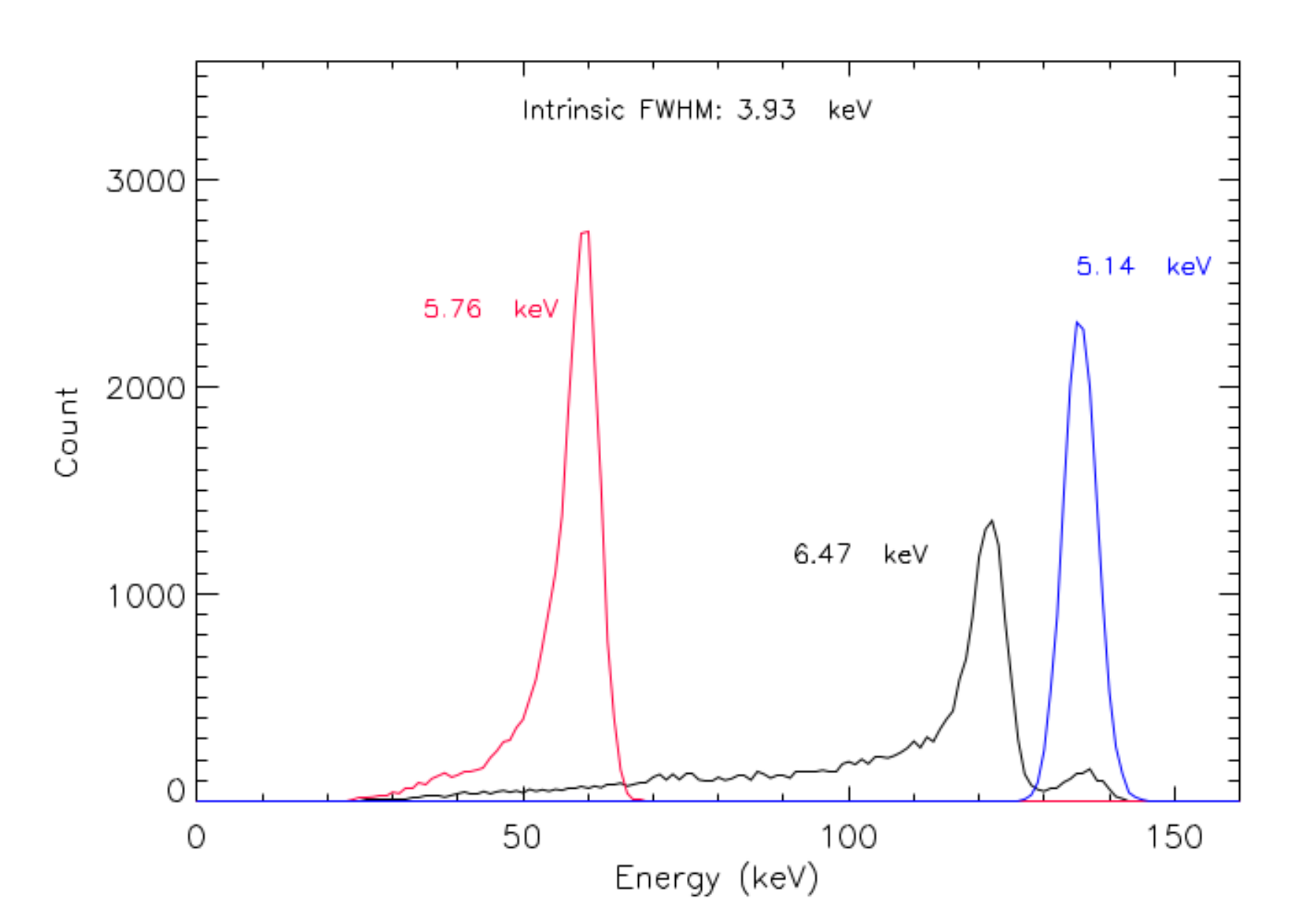}
\caption{Energy spectrum for A3: $^{57}$Co (Black), $^{241}$Am (red) and pulser (blue). A3 width is 0.3 mm.}
\label{fig:FWHM_anode_3}
\end{center}
\end{figure}

\begin{figure}[!htb]
\begin{center}
\includegraphics[width=1.0\textwidth]{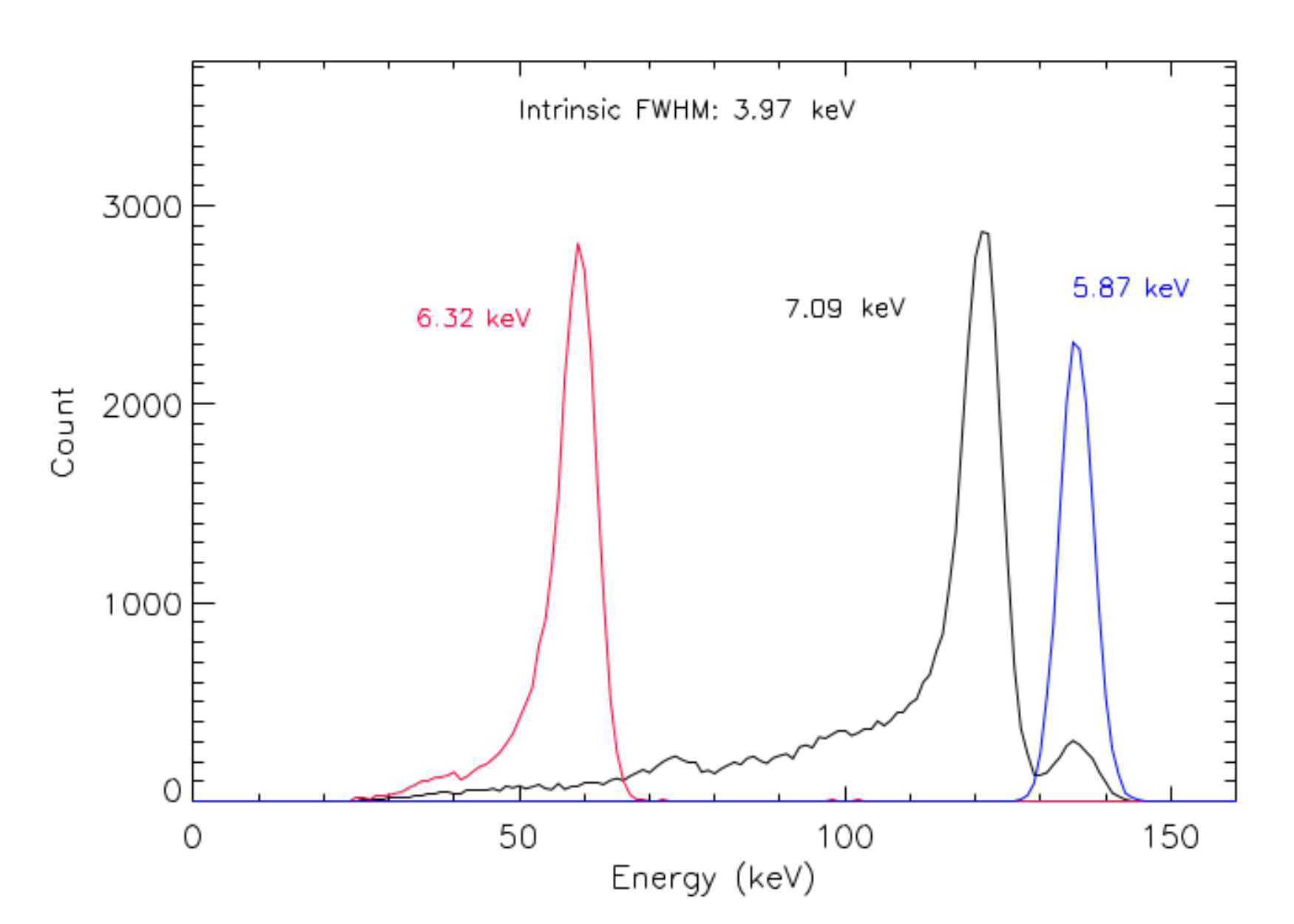}
\caption{Energy spectrum for A12: $^{57}$Co (Black), $^{241}$Am (red) and pulser (blue). A12 width is 0.3 mm.}
\label{fig:FWHM_anode_12}
\end{center}
\end{figure}
For the energy resolution measurements, we considered only the single events. The anodes are at ground, while the SE and the cathode strips are set to -50V and -500V respectively. For each electrode, we determined the FWHM by fitting a Gaussian function to the 122.1 keV peak for $^{57}$Co source, together with the pulser after calibration. One can see the results for A3 and A12 as examples in Figures \ref{fig:FWHM_anode_3} and \ref{fig:FWHM_anode_12}. Since the pulser is directly connected to the input of the charge sensitive preamplifier in the readout electronics, its FWHM provides a measure of the noise level due to the electronical effects only. We determined the intrinsic contribution of the crystal noise by subtracting (in quadrature) the FWHM of pulser's from those of the calibration source line at 122.1 keV. The results show that SE4 group showed the best performance in terms of both the intrinsic and overall energy resolution for the $^{57}$Co peak. On the other hand, SE1, SE2 and SE3 shows slightly better electronic noise performance compared to SE4. All results are summarized in Table \ref{table:widths_2} (see Table \ref{table:widths} for the anode and SE geometry for each set).

\definecolor{Gray}{gray}{0.85}
\definecolor{LightCyan}{rgb}{0.88,1,1}

\definecolor{brightturquoise}{rgb}{0.03, 0.91, 0.87}
\definecolor{Dark_Green}{rgb}{0.0, 0.5, 0.0}
\definecolor{Apricot}{rgb}{0.98, 0.81, 0.69}
\definecolor{brightube}{rgb}{0.82, 0.62, 0.91}
\definecolor{cadmiumred}{rgb}{0.89, 0.0, 0.13}
\begin{table}[!ht]
\centering
	\caption{FWHM values for different SE groups.}
	\label{table:widths_2}  
	\begin{tabular}{|c|c|c|c|c| }
		\hline
        \centering
		\thead{Anode} & \thead{$^{57}$Co \\ FWHM \\ (keV)}& \thead{Pulser \\ FWHM \\ (keV)} &   \thead{Intrinsic \\ $^{57}$Co FWHM \\ (keV)} & \thead{Group FWHM \\ $^{57}$Co (keV)}\\ 
	    \hline
	   	\rowcolor{cadmiumred}
		1 & 7.46  & 5.42  & 5.12*&  \\
		\hline
	\rowcolor{cadmiumred}
		2 & 7.15  &  5.39  & 4.70 &4.32 \\
		\hline
			\rowcolor{cadmiumred}
		3 & 6.47  & 5.14  &3.93 & \\
		\hline
		\rowcolor{brightube}
		4 & 7.13  & 5.48  & 4.57  & \\
		\hline
		\rowcolor{brightube}
		5 & 6.84  & 5.27  & 4.37 & 4.57 \\
		\hline
		\rowcolor{brightube}
		6 & 7.09  & 5.25  & 4.77  &\\
		\hline
	    \rowcolor{Apricot}
		7 & 10.03  & 5.41   &8.44*  &\\
		\hline
		\rowcolor{Apricot}
		8 & 7.61  & 5.58  &5.18  &\\
		\hline
		\rowcolor{Apricot}
		9 & -  & -  & - &4.89 \\
		\hline
		\rowcolor{Apricot}
		10 & 7.36  & 5.75  & 4.6   &\\
		\hline
		\rowcolor{Dark_Green}
		11 & 6.83  & 5.80  & 3.62  &\\
		\hline
		\rowcolor{Dark_Green}
		12 & 7.09  & 5.88  & 3.98 & 3.63\\  
		\hline
		\rowcolor{Dark_Green}
		13 & 6.82  & 5.97  & 3.29  &\\ 
		\hline
		\rowcolor{brightturquoise}
		14 & 7.43  & 6.18  & 4.13  &\\ 
		\hline
		\rowcolor{brightturquoise}
		15 & 8.04  & 6.59  & 4.60 & 4.37\\ 
		\hline
		 \rowcolor{brightturquoise}
		16 & 8.87  & 6.11  & 6.43* & \\
		\multicolumn{5}{l}{*The results are not included in the group FWHM}
		\\
	    \multicolumn{5}{l}{calculation due to having high noise.}
	    \\
	    \multicolumn{5}{l}{**The fit errors for $^{57}$Co and Pulser FWHM values are}
	   \\ \multicolumn{5}{l}{between $\mp{0.002}$ and $\mp{0.006}$}
	\end{tabular}

\end{table}

\subsection{Relation between the energy resolution and bias voltages}
\label{subsec:steerdep}

Applying a potential difference between the SE and anodes pushes electrons to the anodes. High voltage bias will decrease the electron loss due to the recombination and/or trapping on the surface between the strips \cite{Bolotnikov_1999}. Conventionally, 10\% of cathode voltage is applied to the SEs. 

\begin{figure}[!htb]
\begin{center}
\includegraphics[width=1.0\textwidth]{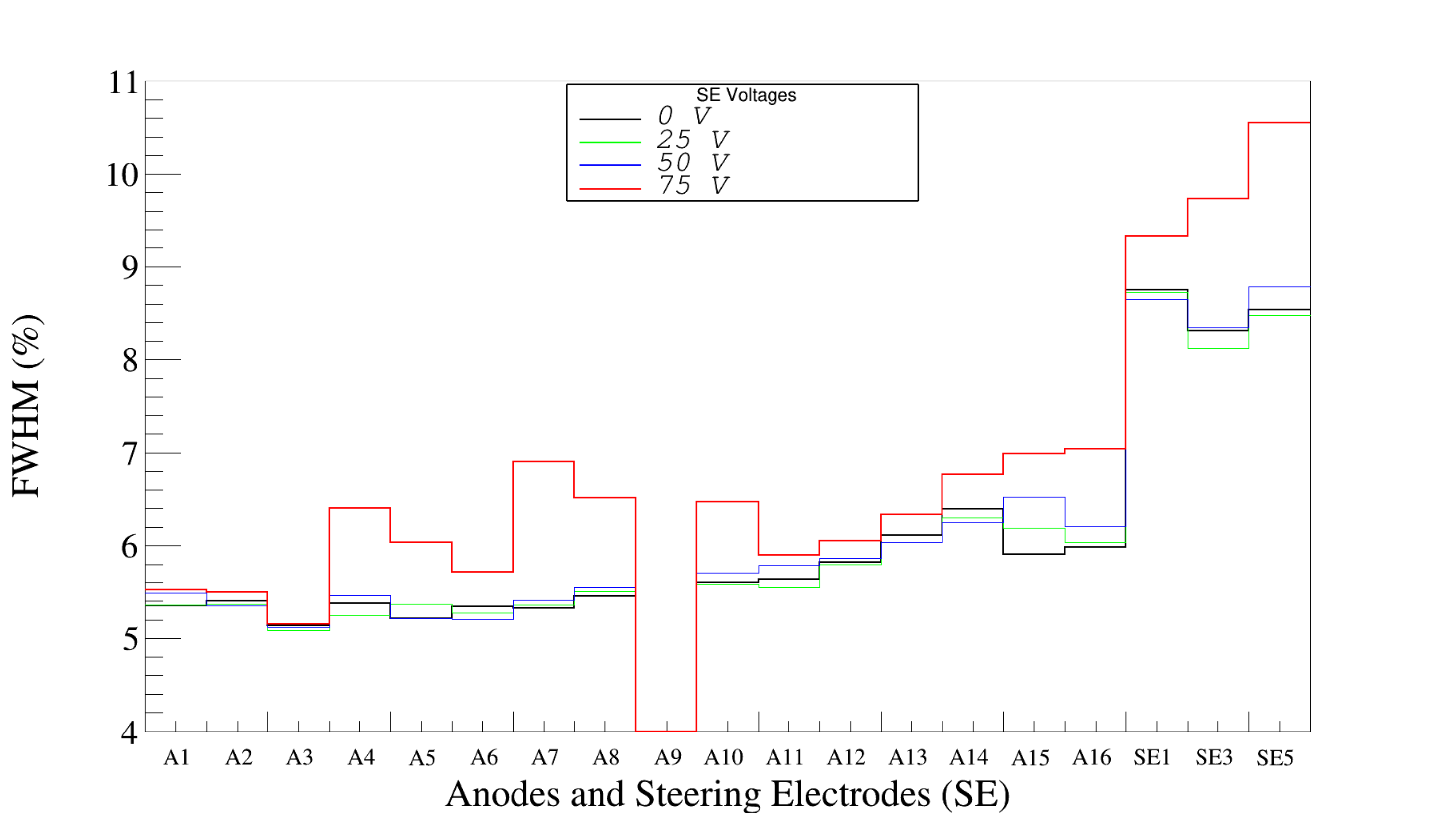}
\caption{FWHM values for each anodes and SE for different SE bias voltages. A denotes anodes. There is no signal from A9.}
\label{fig:en_res_bias}
\end{center}
\end{figure}

On the other hand, high potential difference may increase noise due to an increase in the leakage current  \cite{Luke_2001}. We performed measurements by using a pulser to see the effect of bias between the SE and the anodes to the energy resolution. We used 0 V, -25 V, -50 V and, -75 V values for SE while the cathode voltage was kept constant at -500 V. As one can see in Figure \ref{fig:en_res_bias}, 0 V, -25 V, -50 V potentials yield similar results for the pulser resolution, however, using -75 V potential for the SE results in higher FWHM values. Because of their much larger areas compared to the anode strips, the FWHM values for the SE are higher, as expected. We realized that there is a slope from A5 to A15 in Figure~\ref{fig:en_res_bias}. This slope may be related to the length of the paths in the adapter board circuit (see Figure \ref{fig:systempic}). A1 has the shortest path, while A16 has the longest path. As stated earlier, we could not get any signal from the A9. These results, together with charge sharing studies, implicate that -50 V potential for the SE is optimal for our detector and electronics.

\subsection{Charge Sharing}
\label{sec:ch_sharing}

Charge sharing effect is quite important, since it causes degradation of the spectral performance of detectors \cite{veale_2011}, \cite{Kim_200Conf}. As a result of the expansion in the cloud sizes, the holes may induce charges on several neighboring cathode strips, while the charge induction due to the electron drift may be shared between adjacent SE and anode strips or among neighboring anode strips \cite{Kalemci02}.
 Although the electron cloud expansion is the dominant reason for the charge sharing, Compton scattering and characteristic X-ray emission are the other factors that can contribute as well.

\begin{figure}[H]
\centering
\begin{subfigure}{}
  \centering
  \includegraphics[width=0.8\textwidth]{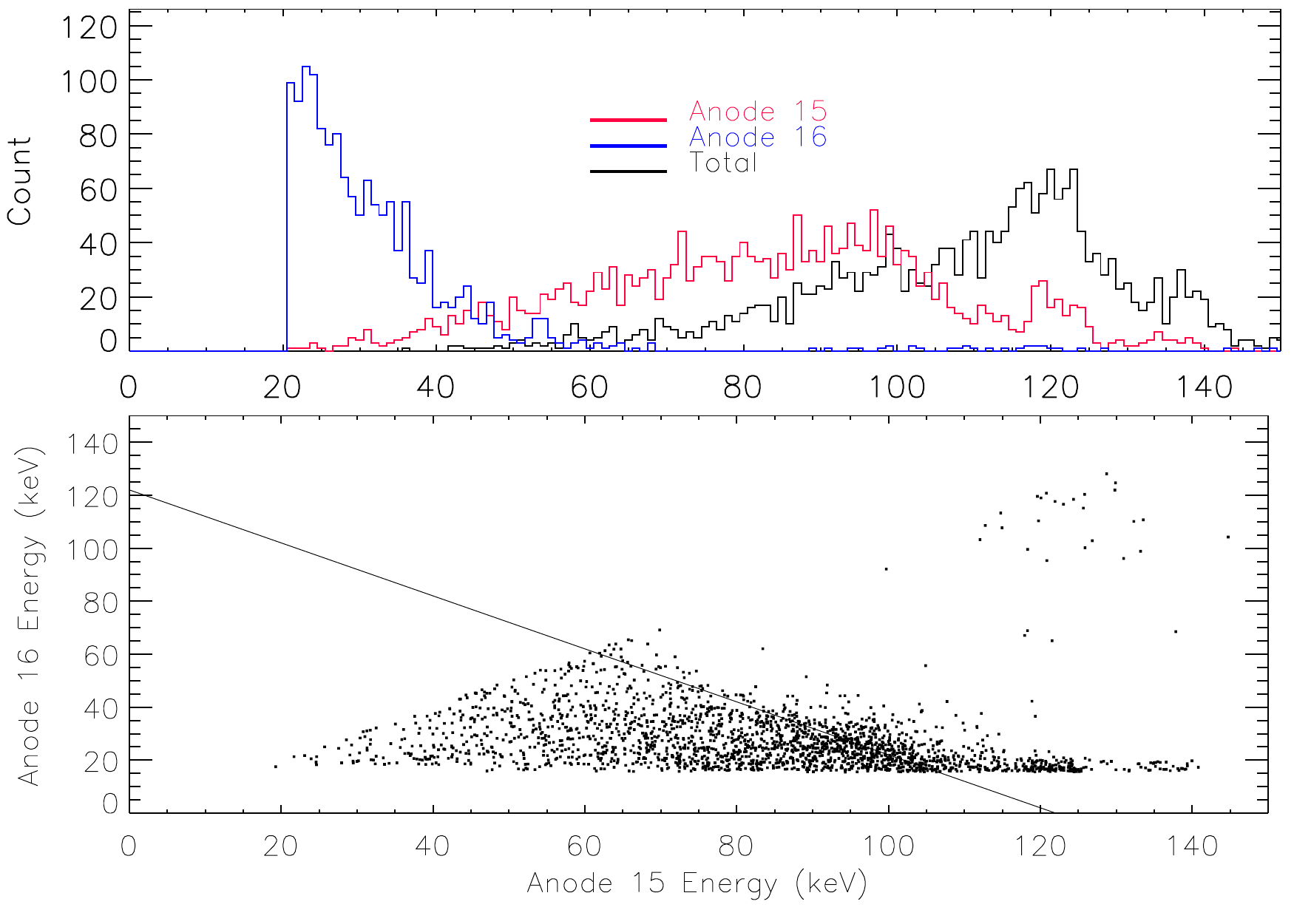}
\end{subfigure}
\begin{subfigure}{}
  \centering
  \includegraphics[width=0.8\textwidth]{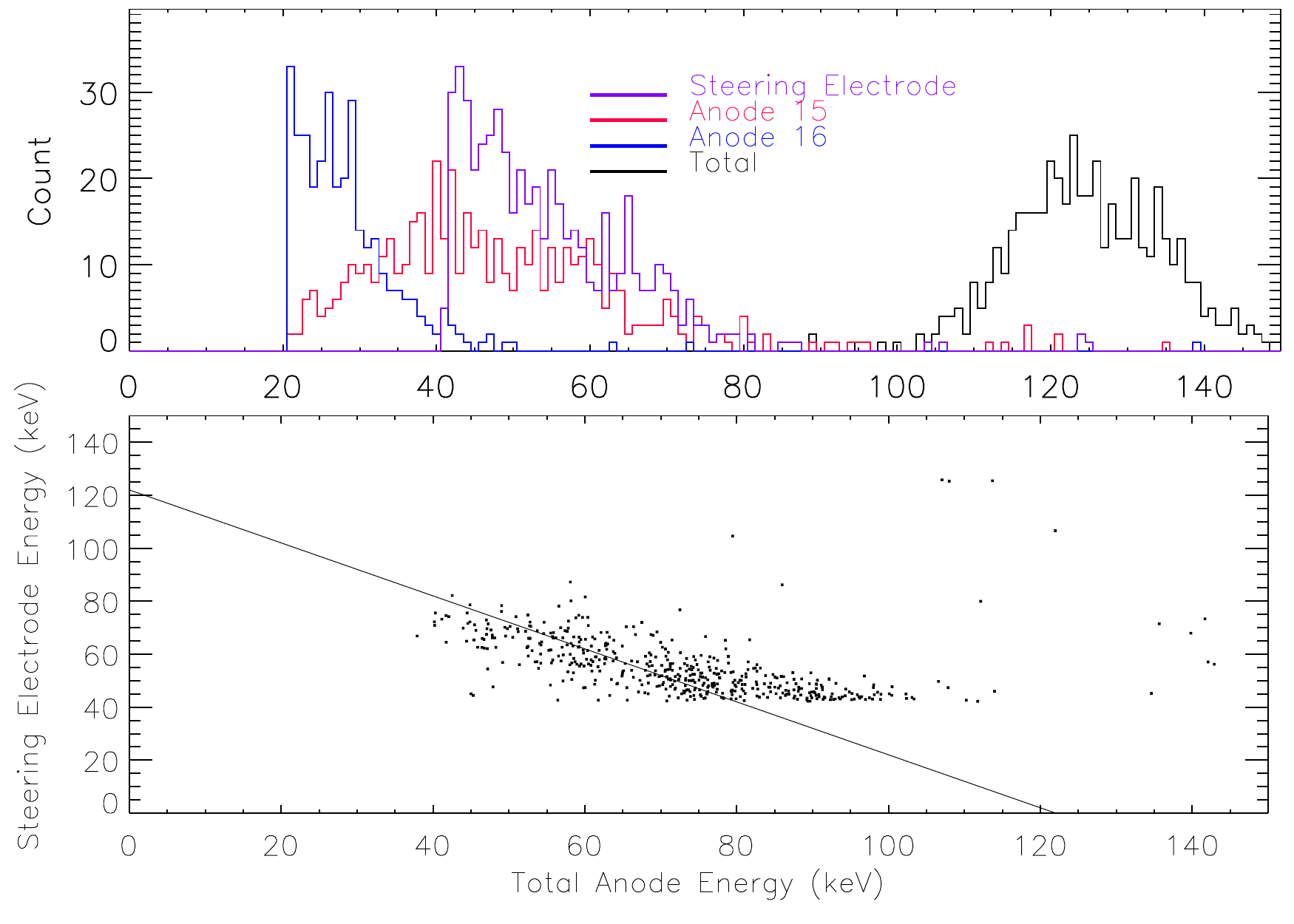}

\end{subfigure}
\caption{Top pair: Shared events between A15 and A16. The events are shared between the anodes, but the SE signal is under threshold. Bottom pair: The events are shared between the anodes and SE5}
\label{fig:plotsharing_14_15}
\end{figure}

We studied the relationship between the charge sharing and the strip geometry. First, the events shared between A15 and A16 and the SE are shown as an example in Figure \ref{fig:plotsharing_14_15}. The figure is divided as two pairs (as top pair and bottom pair) each including an energy spectrum and a graph of strip signals relative to each other. While the results for the shared events between only two anodes (SE signal is less than the threshold) are illustrated in the top pair, the shared events between two anodes and one SE are shown in the bottom pair. Ideally, it is expected in the graph of strip signals that the points should lie in the vicinity of the diagonal line indicating full collection when summed. We calculated the ratio of the number of the shared events between only two neighbouring anode strips and singles registered by the corresponding two anodes with SE signal under threshold (40 keV). The results indicate that the ratios for the anode pairs A2 and A3, A5 and A6, A7 and A8, and A11 and A12 are less than 0.4\%. On the other hand, as one can see from Figure \ref{fig:plotsharing_14_15}, charge sharing effect is visible for A15 and A16 strips which are the thickest ones. For those anode pairs, the ratio of shared events is about 3.5\%. Overall, these results indicate that, except the thickest anodes, the events are not shared significantly between two anodes, however, as discussed below, they can be shared significantly between an anode and a steering anode.

Second, charge sharing between one anode and one SE strip is studied. All of the results are shown in Table \ref{table:table_exp_events_shared}. $N_{Shared}$ is the number of the shared interactions between one anode and one SE. An event which induces signals on a SE and an anode more than the threshold values (40 keV and 20 keV respectively) is considered to be a shared event. $N_{Single}$ is the number of events generating signals only at the anodes. To obtain a meaningful comparison, the energy range for the total signal is set between 90 keV and 150 keV for the shared events as well as for the single events. 

\begin{figure}[H]
\centering
\begin{subfigure}{}
  \centering
  \includegraphics[width=0.8\textwidth]{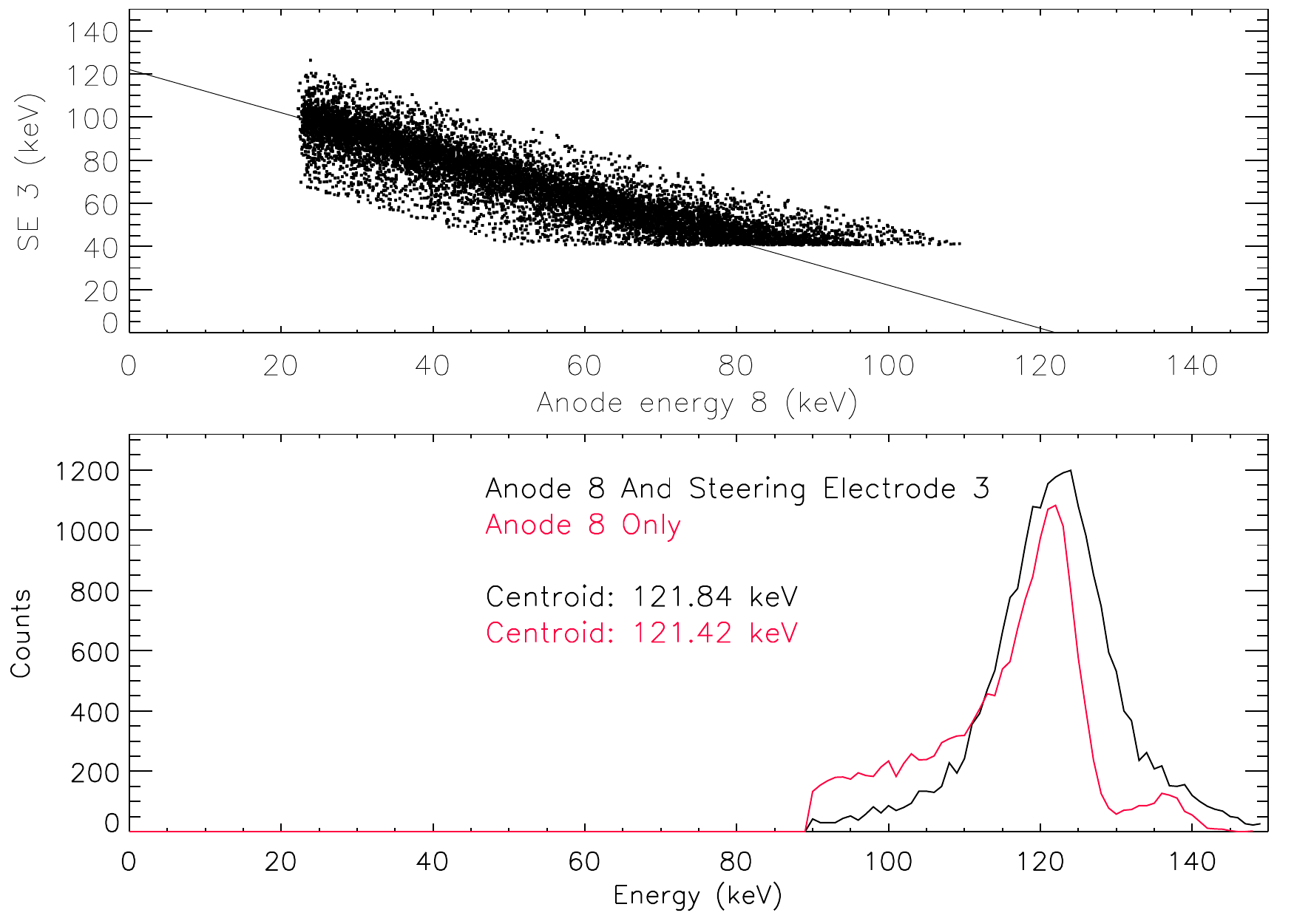}
 \end{subfigure}
\begin{subfigure}{}
  \centering
  \includegraphics[width=0.8\textwidth]{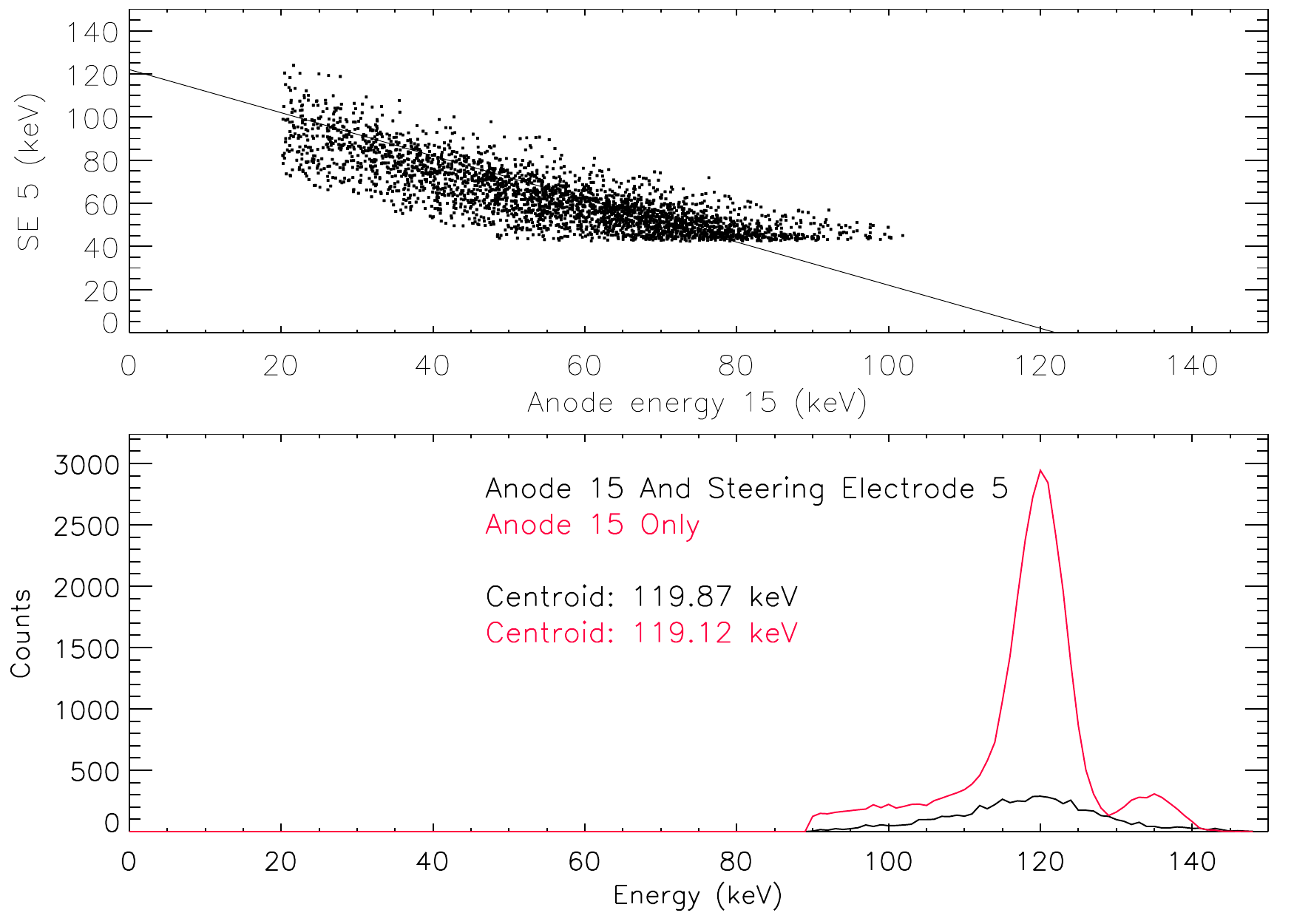}
\end{subfigure}
\caption{Single and shared events for A8 and A15. Strip signals relative to each other and energy spectra for the shared events and the single events for SE3 and A8 (Top pair) and for SE5 and A15 (bottom pair).}
\label{fig:plotsesharing_A8_SE3_A15_SE5_2}
\end{figure}

As the width of the SE increases, the charge sharing effect becomes more dominant. For the third SE, the thickest one, and the fifth SE, the thinnest one, the results are shown in Figure~\ref{fig:plotsesharing_A8_SE3_A15_SE5_2}. While the 122 keV peak is recovered by adding the SE and anode signals, the energy resolution of the combined signal is significantly worse than that of single anode signals.

At this point, it is worth showing the spectra of events collected only at the SEs. As the simulations show (see \S \ref{sec:sim_results}), especially for the thicker SEs, the electric field lines end at the steering electrode, allowing full charge collection at the SE.

\begin{figure}[H]
\begin{center}
\includegraphics[width=1.0\textwidth]{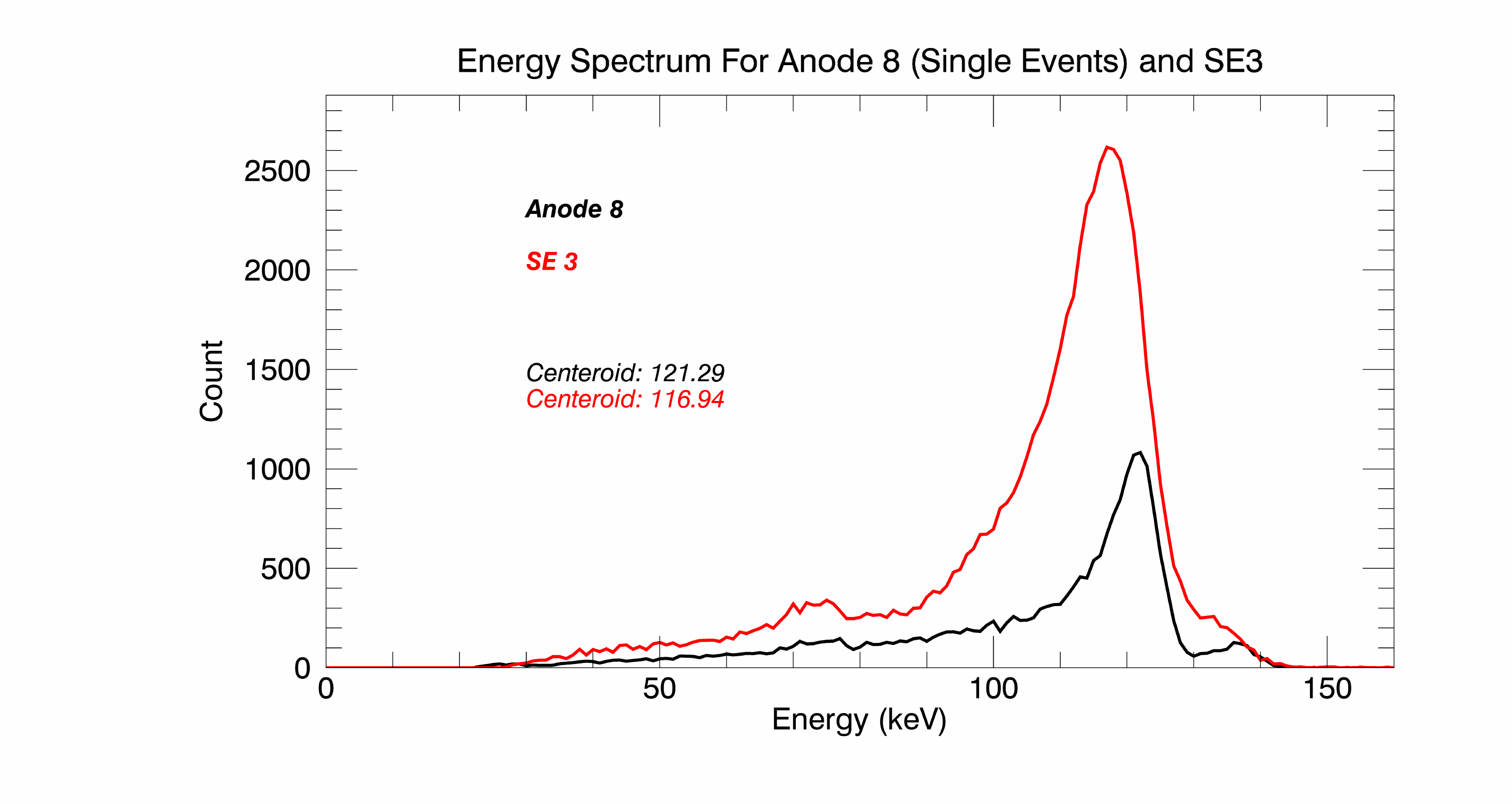}
\caption{Energy spectra for A8 and SE3 singles.}
\label{fig:SE3_only_spec}
\end{center}
\end{figure}

Figure \ref{fig:SE3_only_spec} shows the singles spectra of the largest steering electrode SE3 together with the spectrum of anode numbered 8. In this case, the SE collects much more than the anode. Its energy resolution is notably poor, and the peak energy is less than 122 keV as the large size results in degradation due to hole trapping.

\begin{figure}[H]
\begin{center}
\includegraphics[width=1.0\textwidth]{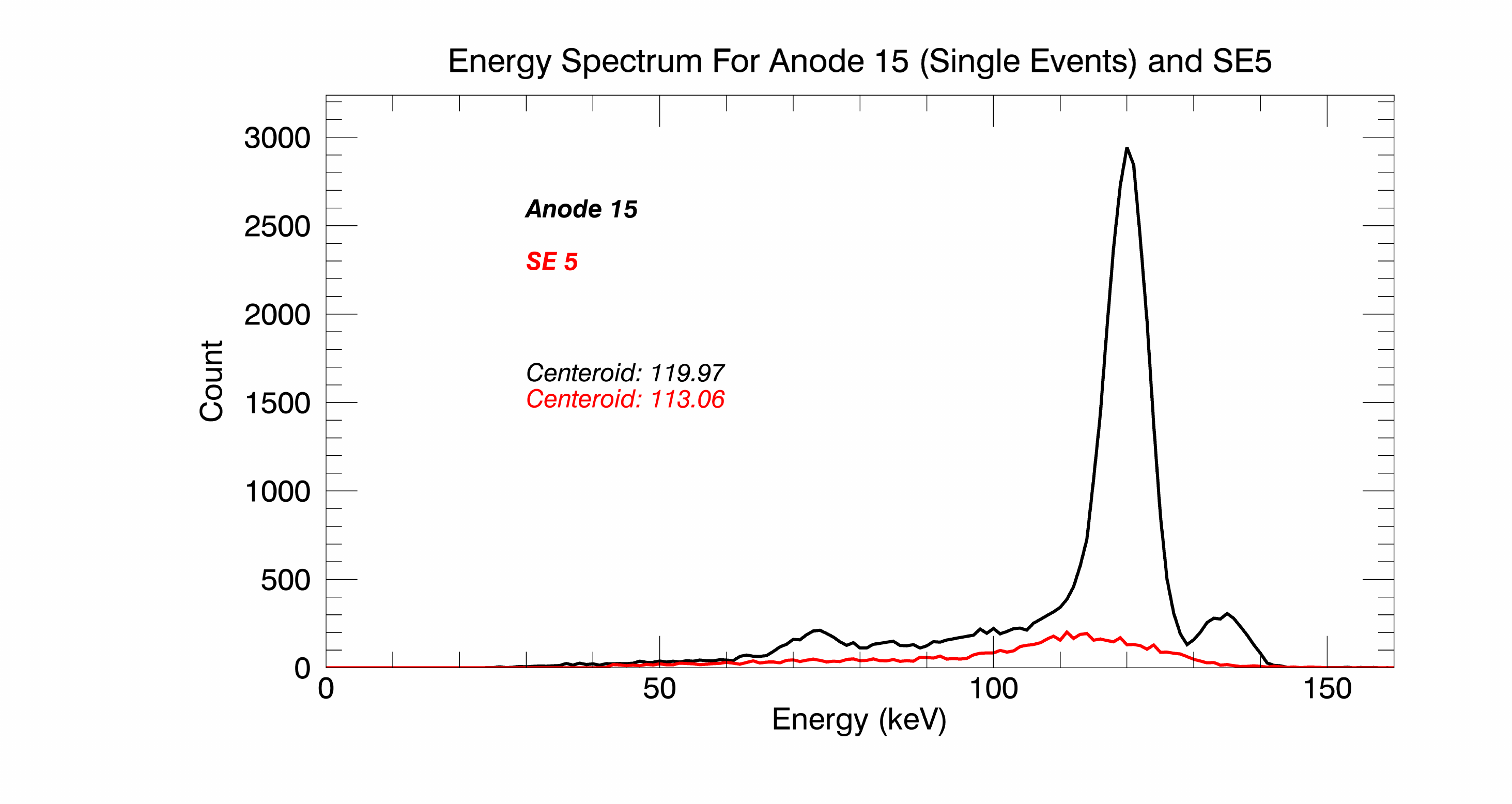}
\caption{Energy spectra for A15 and SE5 singles.}
\label{fig:SE5_only_spec}
\end{center}
\end{figure}

Figure \ref{fig:SE5_only_spec}, on the other hand, shows the singles spectra of the smallest steering electrode SE5 with anode numbered 15 spectrum. In this case, the SE spectrum is negligible and with poor resolution.

\subsection{Simulation results}
\label{sec:sim_results}
The details of the simulation method are given in \S\ref{sec:simulations}. The most important factor that determines the intrinsic resolution of a CdZnTe crystal (excluding electronic noise due to readout electronics) is the charge collection efficiency. The charge collection is reduced by hole trapping as well as by the charges that end up in between the electrodes on the crystal surface. One of the aims of the SE is to minimize the number of charges ending up in this gap. 

\begin{figure}
\begin{center}
\includegraphics[width=1.0\textwidth]{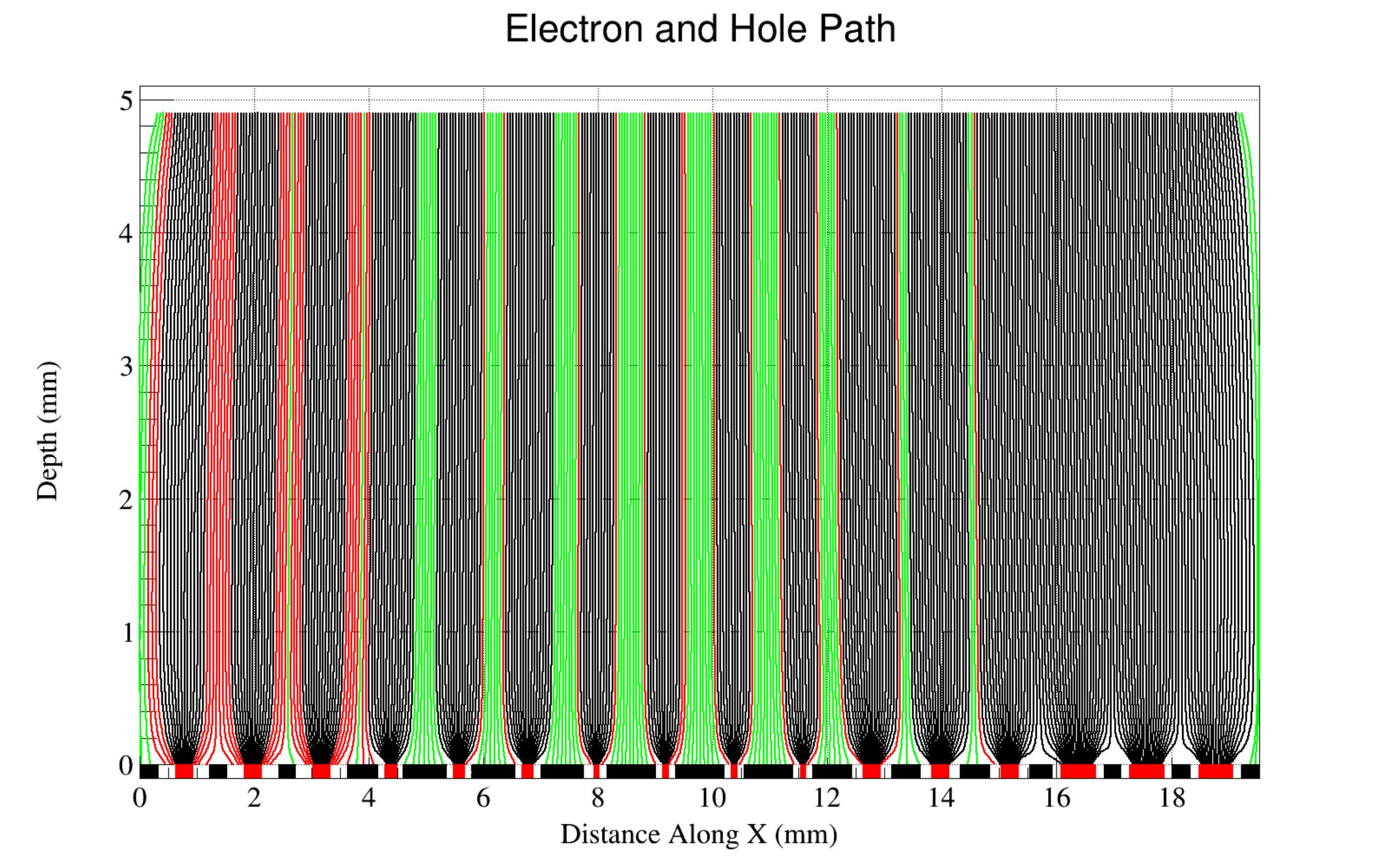}
\caption{Cloud trajectories for interactions close to the cathode surface. Each trajectory is 0.05 mm apart. Black, green and red trajectories indicate clouds ending up in the anodes, steering electrodes and the gaps, respectively. The charges start at the middle of 8th cathode.}
\label{fig:chargedist}
\end{center}
\end{figure}    

First, we have investigated where the electron clouds will end up at the anode side if the interactions take place very close to the cathode side. The simulated potentials are -500 V at cathodes and -50V at steering electrodes while the anodes are at ground. Figure~\ref{fig:chargedist} shows the results. Here each interaction is  0.05 mm apart and the black, green and red trajectories indicate clouds ending up in the anodes, steering electrodes and the gaps, respectively. While 72\% of the electron clouds end up on the anodes, 17\% of electrons arrive at the steering electrodes and 11\% of them are reaching the gaps. The figure immediately shows that, for the nominal 10\% steering electrode to cathode potential ratio, a large fraction of clouds ends up at the steering electrodes that are larger than 0.625 of the pitch (0.75 mm), and therefore we observe the single SE events (Figure \ref{fig:SE3_only_spec}). In theory, it is  possible to reduce events ending up at the steering electrode by increasing the potential difference between the anodes and the steering electrodes, and indeed, simulations indicate that increasing the steering electrode to cathode potential ratio to 20\% results in 95\% of events ending up at anodes. However, at least for our crystal, it was not possible to get any data under such potentials possibly due to surface currents between the anodes and steering electrodes severely affecting the leakage current, and therefore increasing electronic noise significantly. This was discussed in detail in \S\ref{subsec:steerdep}.

\begin{figure}[H]
\begin{center}
\includegraphics[width=1.0\textwidth]{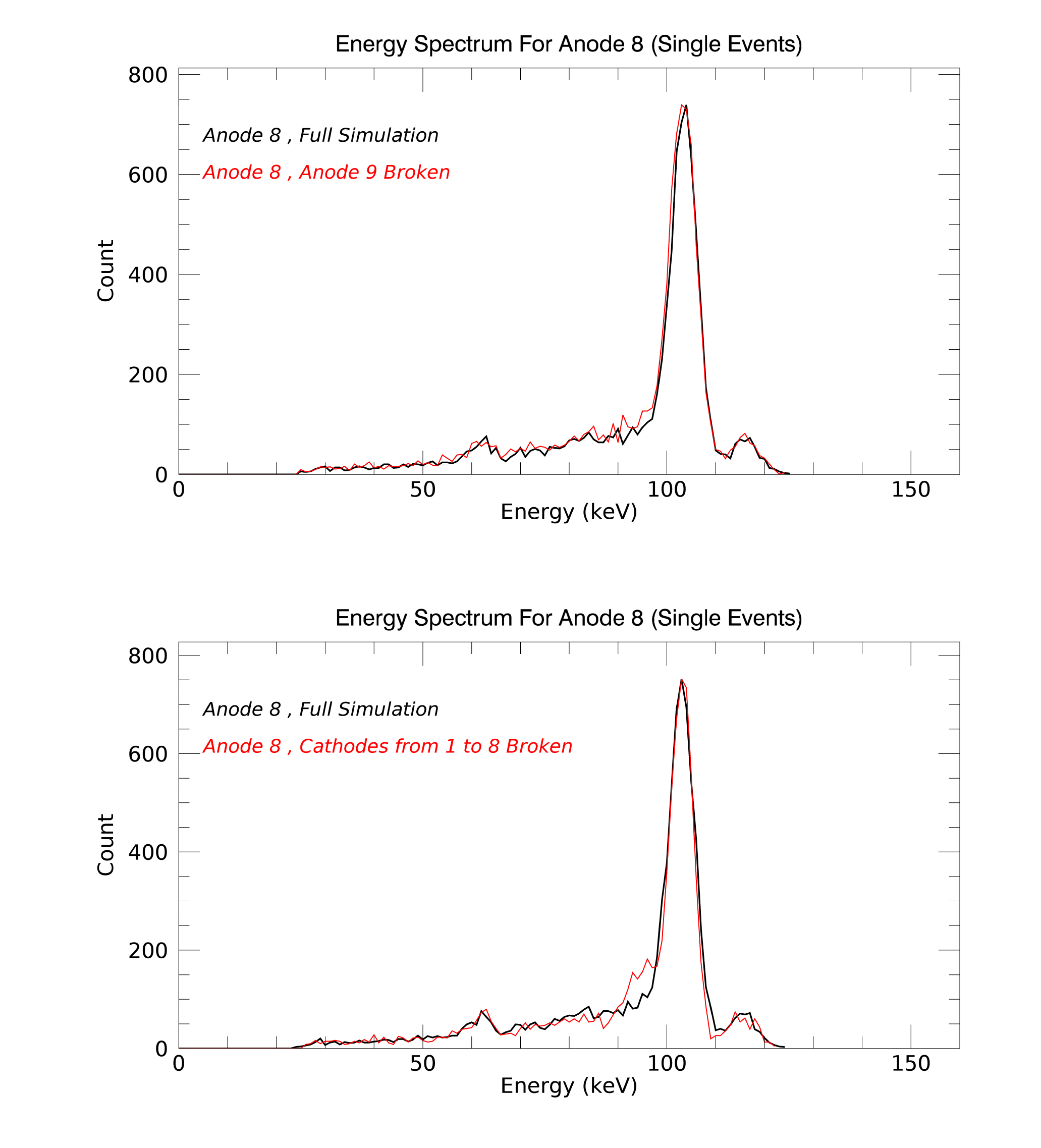}
\caption{Comparison of energy spectra of anode numbered 8 for the cases of anode 9 broken only, cathode 1-8 broken only and the fully functional electrodes. The spectrum for the broken-strip case is normalized.}
\label{fig:broken_strip_comparison_1}
\end{center}
\end{figure}

As mentioned in \S\ref{sec:experiments},
no signal has been received from the cathodes numbered 1-8 and anode numbered 9 due to connection problems and we discarded all events (possibly noise) recorded on cathodes 1-8. We performed additional simulations to determine the effects of non-connected cathode strips and anode numbered 9. Performance of the neighboring anode strip, anode numbered 8, is compared to see the effects. Figure \ref{fig:broken_strip_comparison_1} shows the comparison for the cases of the disconnected anode numbered 9, disconnected cathodes numbered 1-8 and the fully working version of our detector. The spectra are not calibrated to observe whether the cases mentioned above may cause shifts in the energy axis.


 Since anode and cathode strips are orthogonal, charge carriers only in the detector volume under the working cathodes are guided towards to electrodes properly. In the other half (the volume under the non-working cathodes), the charge carriers do not contribute to the signal generation. Therefore, the number of counts for the case of the broken cathode strips is less. In figure \ref{fig:broken_strip_comparison_1}, the spectra are normalized for the comparison purposes. Moreover, one can see that the inoperative cathode strips contribute a little to the low energy tail for both anode numbered 8 and 10, and the disconnected anode has a small low energy tail contribution to the anode numbered 8. However, the simulations indicate no significant change in the energy resolution, and the volume under the properly working cathode strips (from 9 to 16) operates normally.

\subsubsection{Charge Sharing}
\label{sec:sim_charge_sharing}

We  also  examine  the  charge  sharing  phenomenon  in  the  simulations. The results showed that there is no charge sharing between two anodes without a contribution from a steering electrode, even for the thickest ones, anode numbered 14, 15, and 16. The simulation results confirm that the events are not shared among three adjacent electrodes, two anodes and one steering electrode.
Table \ref{table:table_exp_events_shared} shows the charge sharing effect between one anode strip and one steering electrode. One can see from the table that the rates of the shared events for the simulation are less than those for the experiment. However, in agreement with the trend observed experimentally, as the size of the steering electrodes increase, the number of shared events also increase in simulations. For a better comparison, the energy range is set between 90 keV and 150 keV for both the total signal events as well as for the single events. 
\\
\begin{figure}[H]
\centering
\begin{subfigure}{}
  \centering
  \includegraphics[width=0.8\textwidth]{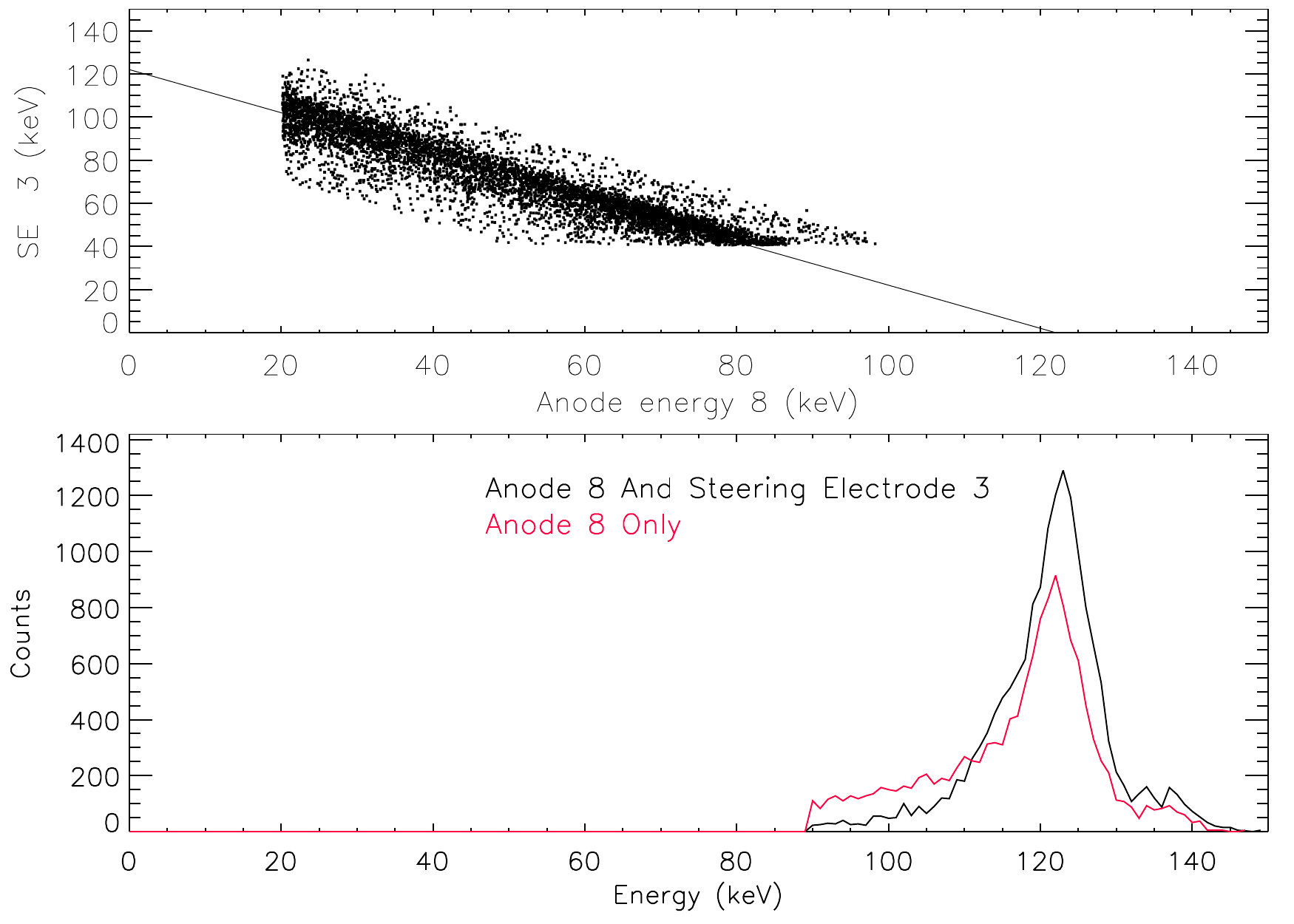}
 \end{subfigure}
\begin{subfigure}{}
  \centering
  \includegraphics[width=0.8\textwidth]{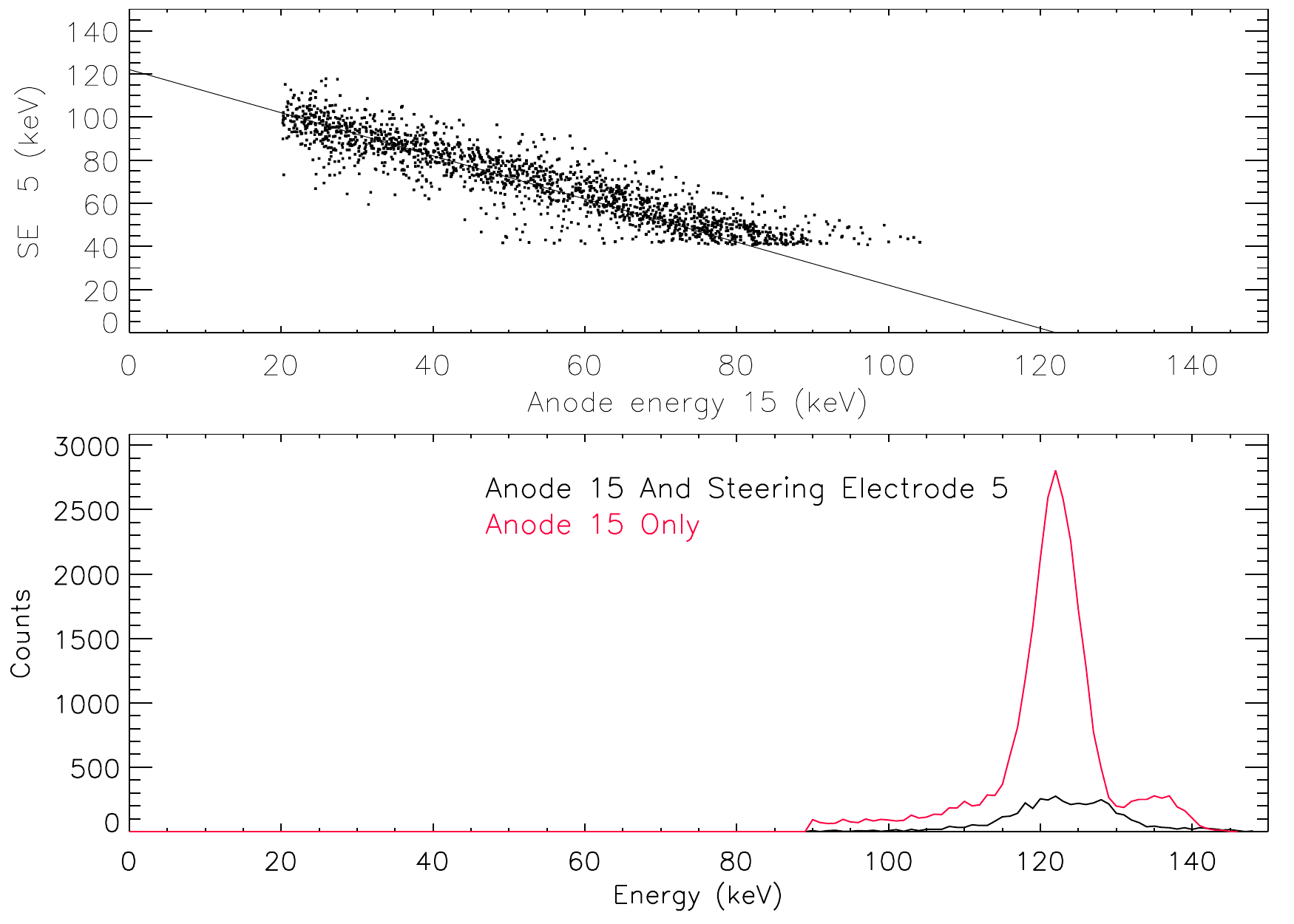}
\end{subfigure}
\caption{Single and shared events for A8 and A15. Strip signals relative to each other and energy spectra for the shared events and the single events for SE3 and A8 (Top pair) and for SE5 and A15 (bottom pair).}
\label{fig:plotsesharing_A8_SE3_A15_SE5_sim}
\end{figure}

For  the  third  SE  and  the  fifth  SE,  the results are shown in Figure \ref{fig:plotsesharing_A8_SE3_A15_SE5_sim}. 

\begin{table}[H]
\centering
\caption{Ratio of the number of shared events ($N_{Shared}$) to number of single events($N_{Single}$) for anodes from each SE group, and the number of single events registered on each anode for the energy range between 90 keV and 150 keV. \newline}

\begin{tabular}{cccccc}
    \hline 
    \\
    EXPERIMENT
    \\
    \hline
    &A3 & A5 & A8 &A12 &A15 \\ 
    &SE1 & SE2 & SE3 & SE4 & SE5 \\
    \hline 
    \\
    \Large{$\frac{N_{Shared}}{N_{Single}}$} & 0.744 $\mp{0.013}$ &  1.065 $\mp{0.017}$ & 1.252 $\mp{0.018}$ & 0.714 $\mp{0.009}$ & 0.189 $\mp{0.003}$ \\ 
    \\
    Collected \\Single Events & 16157 &  14561 & 17076 & 28246 & 32101 \\ 
    \\
    \hline
\end{tabular}
\label{table:table_exp_events_shared}
\end{table}

\begin{table}[H]
\centering
\begin{tabular}{cccccc}
    \hline 
    \\
    SIMULATON
    \\
    \hline
    &A3 & A5 & A8 &A12 &A15 \\ 
    &SE1 & SE2 & SE3 & SE4 & SE5 \\
    \hline 
    \\
   \Large{$\frac{N_{Shared}}{N_{Single}}$} & 0.345 $\mp{0.016}$ &  0.789 $\mp{0.016}$& 1.213 $\mp{0.014}$& 0.451 $\mp{0.012}$ & 0.16 $\mp{0.018}$ \\  
    \\
    Collected \\Single Events & 22128 &  15425 & 13433 & 23125 & 27185 \\ 
    \\
    \hline
\end{tabular}
\label{table:table_sim_events_shared}
\end{table}

\section{Discussion}
\label{sec:dis}
\subsection{Experiment}
\label{subsec:exp_results}
The performance of different anode and steering strip sizes were investigated for an orthogonal-strip design CdZnTe detector.The discussion for the experimental results are given in the following sections in terms of different SE biases, the energy resolution, and the charge sharing effect.
\subsubsection{Steering electrode bias}
\label{subsec:steeringbias}

According to our results, the performance of the system is sensitive to the bias voltage between the SE and the anodes.  Figure~\ref{fig:en_res_bias} shows that the electronic noise was nearly insensitive up to -50V bias to SE. However, at -75V SE bias, the resolution deteriorated significantly. 

Gu et al. (2014)  reported a study on CdZnTe based strip detector with different SE patterns and utilized the RENA-3 ASIC for the data acquisition as we do. However they used the fast trigger mode of RENA optimized for PET imaging \cite{Gu_2014}, while we are using the slow trigger mode optimized for spectroscopy. Their results show that the single events (signals received from one anode and one cathode) are insensitive to the different SE biases (from -40 V to -100 V) when the detector bias was kept at -500 V. They argued that maximizing the SE width and setting a bias of -80 V are important to get a good energy resolution. Moreover, Abbaszadeh et al. (2017) published a study on a PET system using a total of 24 CdZnTe crystals with the cross-strip configuration, also utilizing RENA 3 as the readout ASIC. \cite{Abbaszadeh_2017}. They discussed that better energy resolution can be achieved by increasing the SE bias (between -60V to -80V is the optimal range), and setting the cathode and anode strips to -500 V and 0 V respectively \cite{Abbaszadeh_2017}. But in both studies, the gap widths between the anodes and the SEs were between 0.25 mm and and 0.4 mm. In our study, the largest gap is 0.3 mm (for SE1), while the minimum gap width is 0.125 mm (for SE2 and SE3). In our case, the leakage current from the anodes to the SE may be larger due to the smaller distances between them. This may be the reason that our energy resolution results are poorer compared to \cite{Abbaszadeh_2017} and \cite{Gu_2014} for larger SE biases. 

We emphasize that while the readout electronics is the same, the mode it is operated with, and the crystal manufacturer are different than ours in \cite{Abbaszadeh_2017}. It is clear that a trade-off study needs to be done between the gap size and bias for optimization of CdZnTe detectors during the design stages. 

\subsubsection{Electrode geometry}

In this section how different electrode geometry configurations affect the energy resolution and charge sharing is discussed. 
\\
We analyzed the energy resolution results in two parts: 1) the energy resolution due to readout electronics as indicated by the pulser FWHM, 2) the intrinsic energy resolution of the crystal. 

For the SE1 and SE4 groups, the anode sizes are the same but the SE sizes are bigger for the SE4 (Table \ref{table:widths}). Table~\ref{table:widths_2} illustrates that the overall resolutions are close to one another, while the pulser FWHM values are higher for SE4. For this reason the intrinsic energy resolutions are much better for the SE4 group. One can argue that there are two possible explanations for having higher electronic noise for SE4. One of them is the long connector effect (see Section \ref{subsec:steerdep}) which may create additional noise. The other one is the smaller gap size. Although having smaller gap ensures less electrons to be lost, it is possible that it increases the dark current between the anode and the SE strips as discussed in Section \ref{subsec:steeringbias} increasing the noise.  

The SE1 and SE5 groups were compared to see the effect of the anode size. The SE widths are the same (0.3 mm) for the two groups and the anode sizes for the SE5 are twice as much as the anodes in the SE1 group. The electronic noise is higher for the SE5 group. The larger size of the anodes, meaning larger capacitance, makes them more susceptible to the noise. Also, longer connectors of the SE5 anodes may result in larger pulser FWHM values.  Table~\ref{table:widths_2} shows that the $^{57}$Co overall FWHM values are higher for the last SE group, while the intrinsic FWHM values are nearly the same for the two SE groups. It means that the contribution to the SE5 overall noise mainly came from the electronic component.  

Another pair we could compare are the SE2 and SE3 groups. The gaps between the anodes and the SE are the same (0.125 mm), but the anode sizes for SE2 (0.1 mm) is the half of those of SE3 (0.2 mm). Decreasing the anode size too much and having much larger SE induces more electronic noise, probably due to the increase in the leakage current between the anodes and SE. 

Furthermore, one can see in Table \ref{table:widths_2} that  ${}^{57}$Co FWHM values increase for the anodes close to the edges of the crystal. In the SE1 group, the first and the second anodes yield higher FWHM values than the third one. Also, the last anode, A16, shows notably poorer energy resolution performance compared to the other anodes in the SE5 group. Pulser FWHM values do not have the same tendency. One of the possible explanations for this noise component is that the electric fields on the edges are not uniform which causes more charge carrier loses and hole trapping in the vicinity of the edges.

As an outcome, using large anode strips and small gap sizes increase the electronic noise. Employing larger SE strips induces more electronic noise as well. The SE1 and the SE2 groups give the best performance with respect to low electronic noise.

The intrinsic FWHM values, on the other hand, are related to the defects in the crystals, charge trapping, charge loss in the gaps, strip geometries, and the charge sharing among the neighboring electrodes. First of all, when the results of SE1 and SE4 are compared, we can see the effect of the charge loss to the gaps on the intrinsic resolution. Small gaps ensure less electron losses. The SE4 group which has the smaller gap size performs better intrinsic resolution. Besides, in the SE4 group larger SE sizes keep the weighting potential fields of the anodes more concentrated on a smaller volume (the small pixel effect \citep{Barret}, \citep{Luke_1996}). As a result of that, the anode signals are affected less by the motion of the holes. Further, according to the small pixel effect, larger anode size results in poorer energy resolution performance due to being affected by the hole motion. If we consider only the anode sizes, the SE5 is expected to give worse results compared to those of the SE1 group in terms of the intrinsic resolution. However, in the first group, the anodes are suffering from the charge losses to the gaps which causes them to show poor intrinsic resolution as well. Moreover, employing much larger size SE than the anode strips, for the SE2 and SE3 groups, results in worse intrinsic resolution performances than the other groups. Increasing the size of the SE causes more electron clouds to be shared by more than one electrodes, which degrades the energy resolution. The details of the charge sharing effect will be discussed in the next section.

In conclusion, the energy resolution measurements indicated that the SE4 group showed the best performance in terms of the intrinsic FWHM (see Table~\ref{table:widths_2}). For A8 strip which lies in the thickest SE group, SE3, the intrinsic FWHM became worse. Even the overall FWHM values point out that SE4 is the group with the highest performance. Lastly, while Gu et al. (2014) reported that it is possible to achieve a better energy resolution by increasing the SE sizes \cite{Gu_2014}, our results indicate that, after a point, increasing the SE size will worsen the resolution performance due to other parameters such as leakage currents, electronic noise, and charge sharing effects. 


 Charge sharing is another parameter that affects the performance of the detection system. According to the results reported in \S~\ref{sec:ch_sharing}, charge sharing between two anodes and one SE is not significant except for the last two anodes, A15 and A16. The number of the shared events for those anodes is relatively high because their weighting potentials reach much larger volumes in the detector due to their large widths. For reasonable intrinsic crystal properties, the charge cloud sizes will not be large enough to result in sharing among multiple anodes and a steering electrode.

Charge sharing effect between one anode and one SE is much more common. In fact, Table \ref{table:table_exp_events_shared} and Figure \ref{fig:plotsesharing_A8_SE3_A15_SE5_2} show that the number of shared events between SE2 - A6 and SE3 - A8 pairs are higher than the number of events collected only by the anodes. This result explains why the energy resolution for singles gets worse for the SE2 and SE3 groups. There will be many shared events with SE signals less than the SE threshold registering as singles and producing a tail. 

Directly adding the anode and the SE signals result in significantly poorer energy resolution compared to singles and should not be done unless collecting photons is more important than obtaining good energy resolution. This could be the case for astrophysical applications with faint sources. While it may be possible to compensate for energy losses between electrodes and still obtain a good energy resolution even for shared events in pixellated detectors with small pixel sizes \cite{BUGBY2019}, for the case of strip detectors, especially adding SE signal will always worsen the resolution significantly due to larger intrinsic resolution.


\subsection{Simulations}
We conducted a simulation study for the charge transportation inside the CdZnTe crystals to get a better understanding of the effect of the crystal’s internal properties.
 One issue to mention is related with the electronic noise levels used in the simulatons for the electrodes. We employed a pulser to determine the electronic noise caused by digital circuitry. As it is mentioned in the section \S\ref{sec:en_res_results}, FWHM values measured by the pulser spectra indicated that the readout circuit has 5\% - 6\% contribution to the broadening of the photopeak. However, electronic noise caused by circuits is one component of the total noise. Non uniformity of the defection sites in crystal, distortion of the electric field due to trapped and space charges, non-uniform charge mobilities and the detrapping centers in the detector volume, the bulk and surface conductance \cite{Bolotnikov2002},\cite{Bolotnikov2005}, \cite{Kamieniecki2014} are also important for the intrinsic noise components that severely affects the charge collection efficiency. The simulations indicate that 15\% - 18\% broadening of the photo-peaks gives results consistent with the experiments.

\subsection{Comparison of simulations and experiment}

\begin{figure}[htb!]
\begin{center}
 \includegraphics[width=.9\textwidth]{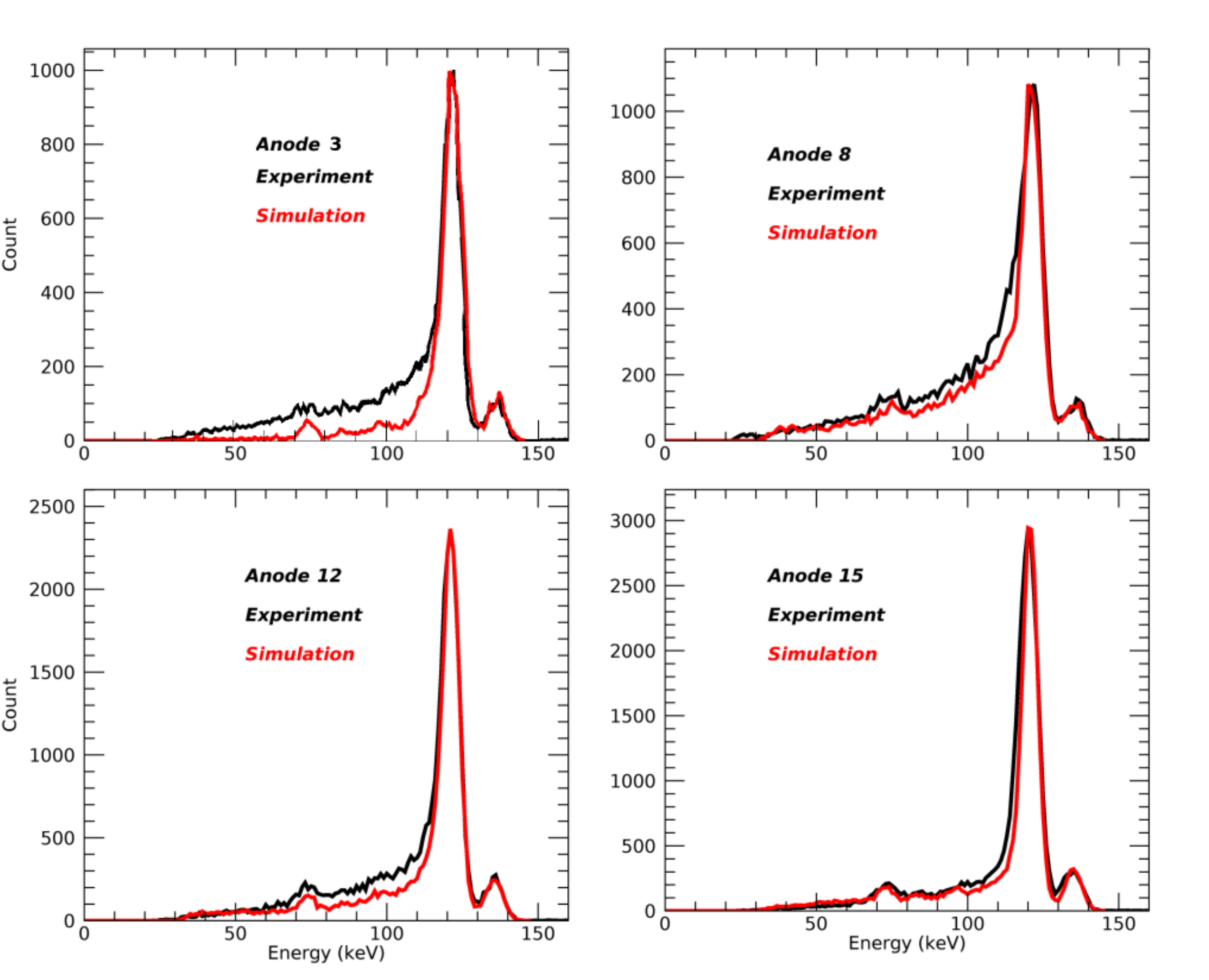}
 \caption{Comparison of experimental and simulation results. Anodes numbered 3, 8, 12, and 15 are used. The simulation results are normalized by using the peak counts from the experiment spectra for the comparative purpose. Parameters for the simulation can be found in the table \ref{table:sim_params}.}
 \label{fig:exp_sim_comparison_all_in_one}
\end{center}
\end{figure}

Figure \ref{fig:exp_sim_comparison_all_in_one} shows a comparison of the simulation and the experimental results. Although there are mismatches especially in the low energy tails of the spectra, the simulations are in good agreement with the experiments. It should be noted that the inconsistency is much more evident for the anode numbered 3.\\

\begin{figure}[htb!]
\begin{center}
 \includegraphics[width=.9\textwidth]{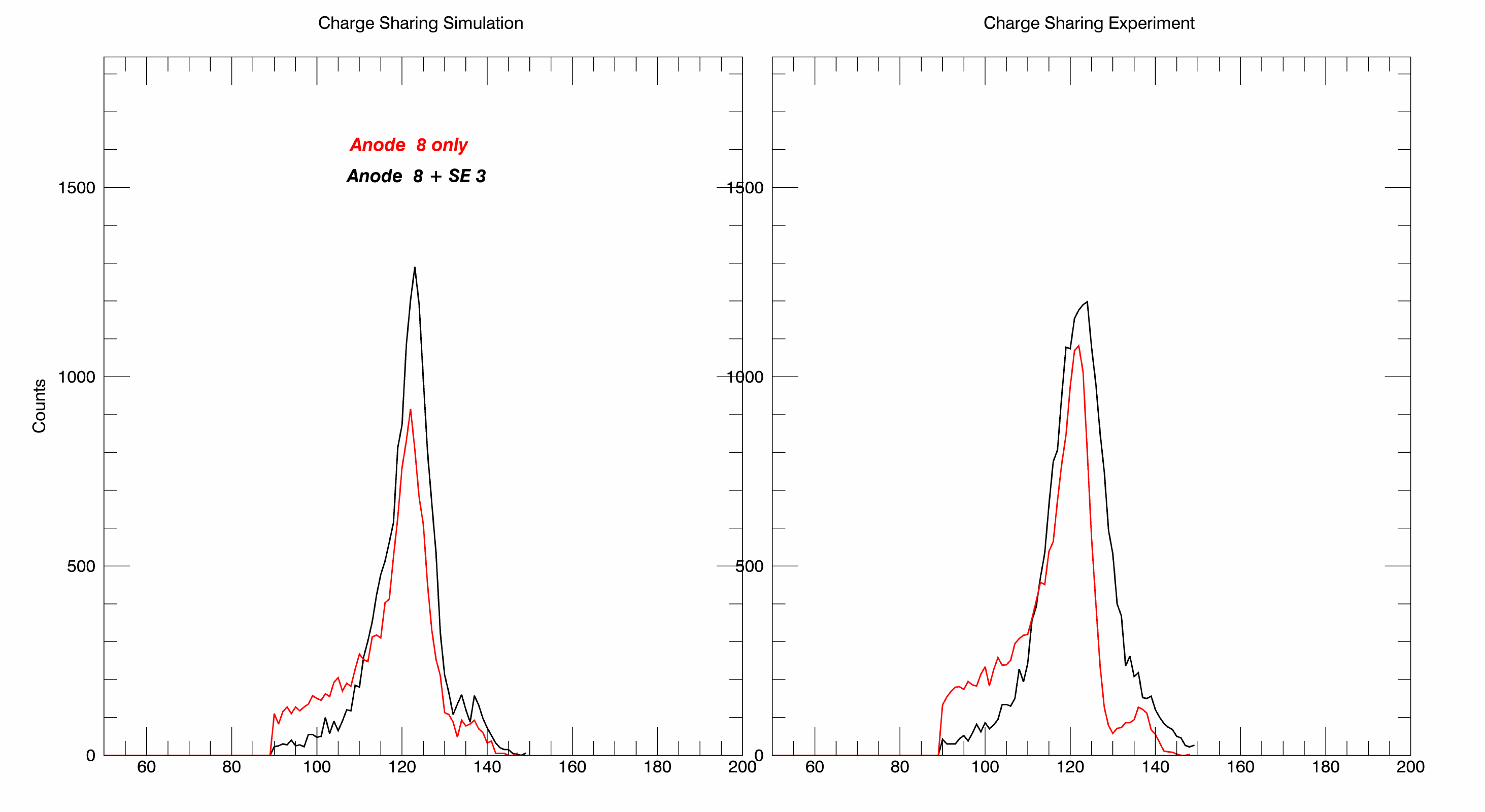}
 \caption{Comparison of the energy spectra for the shared events and the single events for SE3 and A8}
 \label{fig:exp_sim_comparison_CS_A8}
\end{center}
\end{figure}

\begin{figure}[htb!]
\begin{center}
 \includegraphics[width=.9\textwidth]{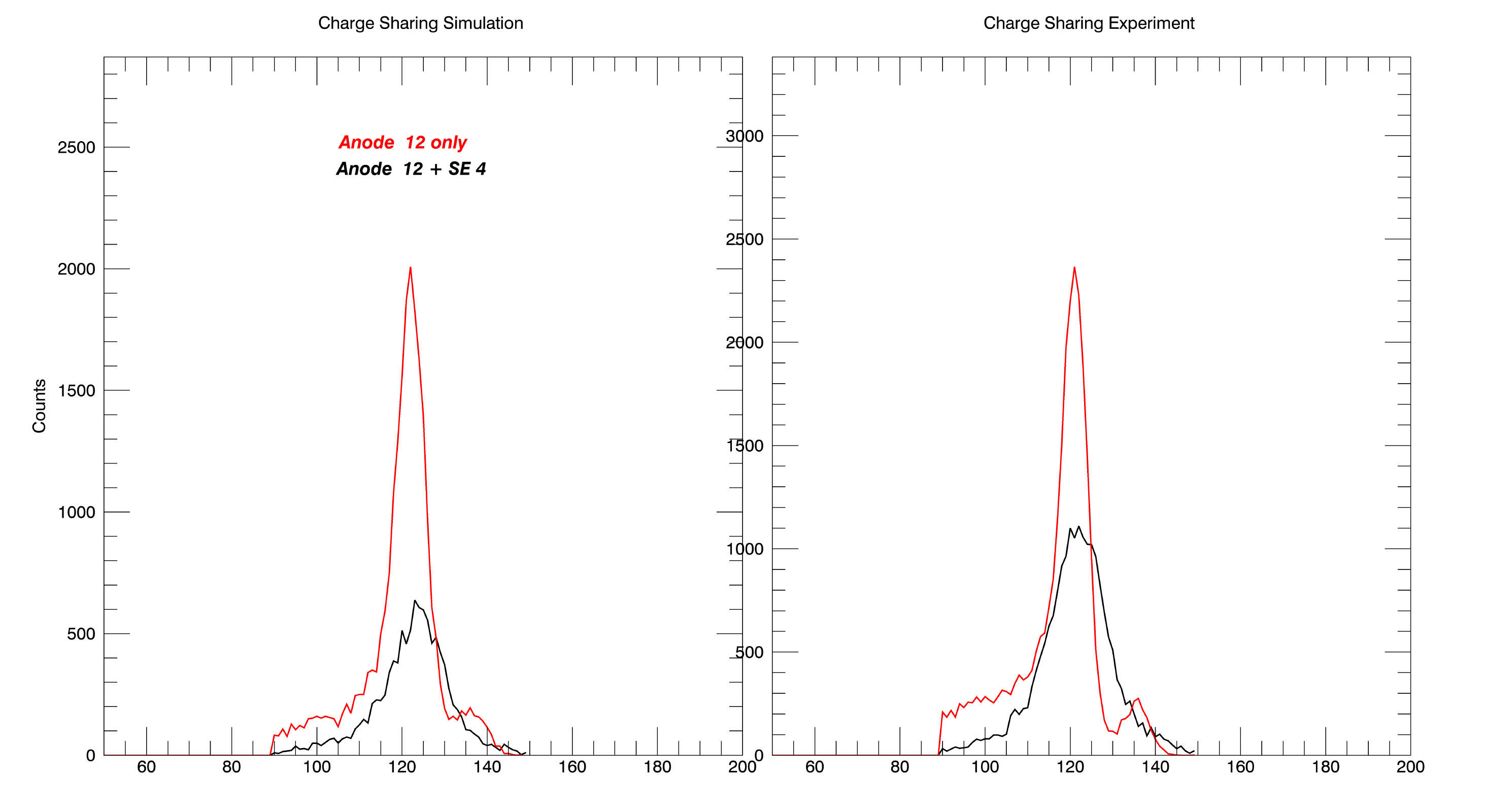}
 \caption{Comparison of the energy spectra for the shared events and the single events for SE4 and A12 }
 \label{fig:exp_sim_comparison_CS_A12}
\end{center}
\end{figure}

Table \ref{table:table_exp_events_shared} indicates that the simulations and experiments exhibit similar
trends with respect to the charge sharing effect. As the size of the steering electrodes increases, the amount of shared events among one anode and neighboring steering electrode scales up. In both works, anode numbered 8, lying in the SE3, causes much more sharing effect than the other anodes, while the thickest anodes, in SE5, yields the least amount of charge sharing. However, as it is illustrated in figures \ref{fig:exp_sim_comparison_CS_A8} and \ref{fig:exp_sim_comparison_CS_A12}, there is a inconsistency in the charge sharing between simulations and experimental results. The numbers of the shared events between anode numbered 3 - SE1 and anode numbered 12 - SE4 are significantly less for the simulations.\\

Surface and bulk conductivity are two important parameters that directly affect the performance of the CZT detector. Dark current and the surface leakage current determine the electronic noise. Also, in the presence of the surface conductivity, electric field lines intersect the gaps between the electrodes, which leads to incomplete charge collection \cite{Bolotnikov2002}, \cite{Bolotnikov2005}. In COMSOL simulations, electric fields are created for an ideal case, where the surface conductivity is ignored. Since the gaps between the anodes and the steering electrode in the first and the fourth group (see Table~\ref{table:widths}) are the large ones, it is possible that we cannot construct full charge collection for those anodes. That, together with the insufficient calculation of the complex electronic noise structure, could be the major reason for the mismatch the charge sharing effect and the low energy tail between the simulation and the experiment. 


\section{Conclusion and future work}
\label{sec:conc}
 We studied the performance of a 5 mm thick orthogonal-strip CdZnTe detector with different electrode sizes with RENA 3b ASIC as the primary readout electronics. The anodes were interspersed with SE strips. The performance of the cross-strip electrode CdZnTe detectors is strongly dependent on the bias voltages, the charge loss to the gaps, the charge sharing effect and the strip sizes. In our system, a SE voltage of -50V is ideal when the anodes are at ground and -500V is applied to the cathode. Employing a higher bias, i.e -75V, degraded the system's performance. In terms of the intrinsic energy resolution and overall resolution, the SE4 group, having 0.3 mm thick anodes and 0.5 mm SE with 0.2 mm gaps, showed the best performance, while the intrinsic resolution was the worst for SE3, the thickest SE group having 0.1 mm anode strips with 0.85 mm SE and 0.125 mm gaps. These results show that increasing SE widths beyond an optimum value worsens the energy resolution performance of the system. In addition to that, as the size of the strips increases, for both SE and the anodes, the electronic noise becomes larger. The small size of the gaps between SE and the anodes increases the electronic noise as well. Finally, the charge sharing between two adjoining anodes scales up as the anode widths increase. Our results show that for anode size less than 0.3 mm the charge sharing between two anodes is quite negligible. The charge sharing between one anode and one SE, on the other hand, is much more frequent than the charge sharing between two anodes. As the SE widths increase, the number of shared events between SE and the anodes increase as well, resulting in poorer energy resolution.
 \\
We also conducted simulations to understand the physics processes inside the CdZnTe crystal. We obtained a good agreement between the simulation and experiment, especially for reconstructing spectra of different anode sets, validating our simulation model. However, our simulation framework has need of better estimation for the intrinsic noise of CdZnTe crystal. The simulation also under-counts the shared events, perhaps due to an idealized electric fields not taking into account surface effects and potential drop due to leakage current between the anodes and the steering electrodes. 
 \\
 A careful optimization study between the electrode sizes and the SE bias is required to obtain the best performance from a cross-strip design. As a future work, we are working to improve our simulation framework to estimate more realistic charge transport and collections inside the CdZnTe crystal and charge sharing which will help the optimization process for future designs.
\section*{Acknowledgment}
This work has been supported by TUBITAK Grants 108T595 and 116F151. EK acknowledges the contributions of \"Ozge (Amutkan) Aygar, Ali Atasever and Onur Akbal in conducting some of the experiments, and \.Irfan Kuvvetli for his contributions to setting up the entire lab.


\begin{thebibliography}{36}
	\expandafter\ifx\csname natexlab\endcsname\relax\def\natexlab#1{#1}\fi
	\providecommand{\url}[1]{\texttt{#1}}
	\providecommand{\href}[2]{#2}
	\providecommand{\path}[1]{#1}
	\providecommand{\DOIprefix}{doi:}
	\providecommand{\ArXivprefix}{arXiv:}
	\providecommand{\URLprefix}{URL: }
	\providecommand{\Pubmedprefix}{pmid:}
	\providecommand{\doi}[1]{\href{http://dx.doi.org/#1}{\path{#1}}}
	\providecommand{\Pubmed}[1]{\href{pmid:#1}{\path{#1}}}
	\providecommand{\bibinfo}[2]{#2}
	\ifx\xfnm\relax \def\xfnm[#1]{\unskip,\space#1}\fi
	\bibitem[{{Kalemci} and {Matteson}(2002)}]{Kalemci02}
	\bibinfo{author}{E.~{Kalemci}}, \bibinfo{author}{J.~L. {Matteson}},
	\newblock \bibinfo{title}{{Investigation of charge sharing among electrode
			strips for a CdZnTe detector}},
	\newblock \bibinfo{journal}{Nuclear Instruments and Methods in Physics Research
		A} \bibinfo{volume}{478} (\bibinfo{year}{2002}) \bibinfo{pages}{527--537}.
	\DOIprefix\doi{10.1016/S0168-9002(01)00892-0}.
	\bibitem[{{Feng Zhang} et~al.(2005){Feng Zhang}, {Zhong He}, {Xu}, and
		{Meng}}]{Zhang_2005}
	\bibinfo{author}{{Feng Zhang}}, \bibinfo{author}{{Zhong He}},
	\bibinfo{author}{D.~{Xu}}, \bibinfo{author}{L.~J. {Meng}},
	\newblock \bibinfo{title}{Feasibility study of using two 3-d position sensitive
		czt detectors for small animal pet},
	\newblock in: \bibinfo{booktitle}{IEEE Nuclear Science Symposium Conference
		Record, 2005}, volume~\bibinfo{volume}{3}, \bibinfo{year}{2005}, pp.
	\bibinfo{pages}{4 pp.--1585}. \DOIprefix\doi{10.1109/NSSMIC.2005.1596621}.
	\bibitem[{Kuvvetli et~al.(2014)Kuvvetli, Budtz-Jørgensen, Zappettini,
		Zambelli, Benassi, Kalemci, Caroli, Stephen, and Auricchio}]{kuvvetli_2014}
	\bibinfo{author}{I.~Kuvvetli}, \bibinfo{author}{C.~Budtz-Jørgensen},
	\bibinfo{author}{A.~Zappettini}, \bibinfo{author}{N.~Zambelli},
	\bibinfo{author}{G.~Benassi}, \bibinfo{author}{E.~Kalemci},
	\bibinfo{author}{E.~Caroli}, \bibinfo{author}{J.~B. Stephen},
	\bibinfo{author}{N.~Auricchio},
	\newblock \bibinfo{title}{{A 3D CZT high resolution detector for x- and
			gamma-ray astronomy}},
	\newblock in: \bibinfo{editor}{A.~D. Holland}, \bibinfo{editor}{J.~Beletic}
	(Eds.), \bibinfo{booktitle}{High Energy, Optical, and Infrared Detectors for
		Astronomy VI}, volume \bibinfo{volume}{9154},
	\bibinfo{organization}{International Society for Optics and Photonics},
	\bibinfo{publisher}{SPIE}, \bibinfo{year}{2014}, pp. \bibinfo{pages}{272 --
		281}. \URLprefix \url{https://doi.org/10.1117/12.2055119}.
	\DOIprefix\doi{10.1117/12.2055119}.
	\bibitem[{Abbaszadeh et~al.(2016)Abbaszadeh, Gu, Reynolds, and
		Levin}]{Abbaszadeh_2016}
	\bibinfo{author}{S.~Abbaszadeh}, \bibinfo{author}{Y.~Gu},
	\bibinfo{author}{P.~D. Reynolds}, \bibinfo{author}{C.~S. Levin},
	\newblock \bibinfo{title}{Characterization of a sub-assembly of 3d position
		sensitive cadmium zinc telluride detectors and electronics from a
		sub-millimeter resolution {PET} system},
	\newblock \bibinfo{journal}{Physics in Medicine and Biology}
	\bibinfo{volume}{61} (\bibinfo{year}{2016}) \bibinfo{pages}{6733--6753}.
	\URLprefix \url{https://doi.org/10.1088/0031-9155/61/18/6733}.
	\DOIprefix\doi{10.1088/0031-9155/61/18/6733}.
	\bibitem[{Zheng et~al.(2016)Zheng, Cheng, Deen, and Peng}]{Zheng_2016}
	\bibinfo{author}{X.~Zheng}, \bibinfo{author}{Z.~Cheng}, \bibinfo{author}{M.~J.
		Deen}, \bibinfo{author}{H.~Peng},
	\newblock \bibinfo{title}{Improving the spatial resolution in czt detectors
		using charge sharing effect and transient signal analysis: Simulation study},
	\newblock \bibinfo{journal}{Nuclear Instruments and Methods in Physics Research
		Section A: Accelerators, Spectrometers, Detectors and Associated Equipment}
	\bibinfo{volume}{808} (\bibinfo{year}{2016}) \bibinfo{pages}{60 -- 70}.
	\URLprefix
	\url{http://www.sciencedirect.com/science/article/pii/S0168900215013686}.
	\DOIprefix\doi{https://doi.org/10.1016/j.nima.2015.11.006}.
	\bibitem[{Meng and He(2005)}]{Meng_2005}
	\bibinfo{author}{L.~Meng}, \bibinfo{author}{Z.~He},
	\newblock \bibinfo{title}{Exploring the limiting timing resolution for large
		volume czt detectors with waveform analysis},
	\newblock \bibinfo{journal}{Nuclear Instruments and Methods in Physics Research
		Section A: Accelerators, Spectrometers, Detectors and Associated Equipment}
	\bibinfo{volume}{550} (\bibinfo{year}{2005}) \bibinfo{pages}{435 -- 445}.
	\URLprefix
	\url{http://www.sciencedirect.com/science/article/pii/S0168900205012295}.
	\DOIprefix\doi{https://doi.org/10.1016/j.nima.2005.04.076}.
	\bibitem[{Zhao et~al.(2012)Zhao, Ouyang, Xu, Han, Zhang, Wang, Zha, and
		Ouyang}]{Zhao_2012}
	\bibinfo{author}{X.~C. Zhao}, \bibinfo{author}{X.~P. Ouyang},
	\bibinfo{author}{Y.~D. Xu}, \bibinfo{author}{H.~T. Han},
	\bibinfo{author}{Z.~C. Zhang}, \bibinfo{author}{T.~Wang},
	\bibinfo{author}{G.~Q. Zha}, \bibinfo{author}{X.~Ouyang},
	\newblock \bibinfo{title}{Time response of cd0.9zn0.1te crystals under
		transient and pulsed irradiation},
	\newblock \bibinfo{journal}{AIP Advances} \bibinfo{volume}{2}
	(\bibinfo{year}{2012}) \bibinfo{pages}{012162}.
	\DOIprefix\doi{10.1063/1.3693970}.
	\bibitem[{{Kalemci} et~al.(2013){Kalemci}, {Ümit}, and {Aslan}}]{Kalemci13}
	\bibinfo{author}{E.~{Kalemci}}, \bibinfo{author}{E.~{Ümit}},
	\bibinfo{author}{R.~{Aslan}},
	\newblock \bibinfo{title}{X-ray detector on 2u cubesat beeaglesat of qb50},
	\newblock in: \bibinfo{booktitle}{2013 6th International Conference on Recent
		Advances in Space Technologies (RAST)}, \bibinfo{year}{2013}, pp.
	\bibinfo{pages}{899--902}. \DOIprefix\doi{10.1109/RAST.2013.6581341}.
	\bibitem[{{Kormoll} et~al.(2011){Kormoll}, {Fiedler}, {Golnik}, {Heidel},
		{Kempe}, {Schoene}, {Sobiella}, {Zuber}, and {Enghardt}}]{Kormoll_2011}
	\bibinfo{author}{T.~{Kormoll}}, \bibinfo{author}{F.~{Fiedler}},
	\bibinfo{author}{C.~{Golnik}}, \bibinfo{author}{K.~{Heidel}},
	\bibinfo{author}{M.~{Kempe}}, \bibinfo{author}{S.~{Schoene}},
	\bibinfo{author}{M.~{Sobiella}}, \bibinfo{author}{K.~{Zuber}},
	\bibinfo{author}{W.~{Enghardt}},
	\newblock \bibinfo{title}{A prototype compton camera for in-vivo dosimetry of
		ion beam cancer irradiation},
	\newblock in: \bibinfo{booktitle}{2011 IEEE Nuclear Science Symposium
		Conference Record}, \bibinfo{year}{2011}, pp. \bibinfo{pages}{3484--3487}.
	\DOIprefix\doi{10.1109/NSSMIC.2011.6152639}.
	\bibitem[{Hueso-Gonz{\'{a}}lez et~al.(2014)Hueso-Gonz{\'{a}}lez, Golnik,
		Berthel, Dreyer, Enghardt, Fiedler, Heidel, Kormoll, Rohling, Schöne,
		Schwengner, Wagner, and Pausch}]{Gonzalez_2014}
	\bibinfo{author}{F.~Hueso-Gonz{\'{a}}lez}, \bibinfo{author}{C.~Golnik},
	\bibinfo{author}{M.~Berthel}, \bibinfo{author}{A.~Dreyer},
	\bibinfo{author}{W.~Enghardt}, \bibinfo{author}{F.~Fiedler},
	\bibinfo{author}{K.~Heidel}, \bibinfo{author}{T.~Kormoll},
	\bibinfo{author}{H.~Rohling}, \bibinfo{author}{S.~Schöne},
	\bibinfo{author}{R.~Schwengner}, \bibinfo{author}{A.~Wagner},
	\bibinfo{author}{G.~Pausch},
	\newblock \bibinfo{title}{Test of compton camera components for prompt gamma
		imaging at the {ELBE} bremsstrahlung beam},
	\newblock \bibinfo{journal}{Journal of Instrumentation} \bibinfo{volume}{9}
	(\bibinfo{year}{2014}) \bibinfo{pages}{P05002--P05002}. \URLprefix
	\url{https://doi.org/10.1088/1748-0221/9/05/p05002}.
	\DOIprefix\doi{10.1088/1748-0221/9/05/p05002}.
	\bibitem[{Owe et~al.(2019)Owe, Kuvvetli, Budtz-J{\O}rgensen, and
		Zoglauer}]{Owe_2019}
	\bibinfo{author}{S.~H. Owe}, \bibinfo{author}{I.~Kuvvetli},
	\bibinfo{author}{C.~Budtz-J{\O}rgensen}, \bibinfo{author}{A.~Zoglauer},
	\newblock \bibinfo{title}{Evaluation of a compton camera concept using the 3d
		{CdZnTe} drift strip detectors},
	\newblock \bibinfo{journal}{Journal of Instrumentation} \bibinfo{volume}{14}
	(\bibinfo{year}{2019}) \bibinfo{pages}{C01020--C01020}. \URLprefix
	\url{https://doi.org/10.1088/1748-0221/14/01/c01020}.
	\DOIprefix\doi{10.1088/1748-0221/14/01/c01020}.
	\bibitem[{Abbene et~al.(2020)Abbene, Gerardi, Principato, Buttacavoli, Altieri,
		Protti, Tomarchio, Del~Sordo, Auricchio, Bettelli, Amad{\`{e}}, Zanettini,
		Zappettini, and Caroli}]{Abbene_2020}
	\bibinfo{author}{L.~Abbene}, \bibinfo{author}{G.~Gerardi},
	\bibinfo{author}{F.~Principato}, \bibinfo{author}{A.~Buttacavoli},
	\bibinfo{author}{S.~Altieri}, \bibinfo{author}{N.~Protti},
	\bibinfo{author}{E.~Tomarchio}, \bibinfo{author}{S.~Del~Sordo},
	\bibinfo{author}{N.~Auricchio}, \bibinfo{author}{M.~Bettelli},
	\bibinfo{author}{N.~S. Amad{\`{e}}}, \bibinfo{author}{S.~Zanettini},
	\bibinfo{author}{A.~Zappettini}, \bibinfo{author}{E.~Caroli},
	\newblock \bibinfo{title}{{Recent advances in the development of
			high-resolution 3D cadmium{--}zinc{--}telluride drift strip detectors}},
	\newblock \bibinfo{journal}{Journal of Synchrotron Radiation}
	\bibinfo{volume}{27} (\bibinfo{year}{2020}) \bibinfo{pages}{1564--1576}.
	\URLprefix \url{https://doi.org/10.1107/S1600577520010747}.
	\DOIprefix\doi{10.1107/S1600577520010747}.
	\bibitem[{{Matteson} et~al.(2008){Matteson}, {Gu}, {Skelton}, {Deal},
		{Stephan}, {Duttweiler}, {Huszar}, {Gasaway}, and {Levin}}]{Matteson_2008}
	\bibinfo{author}{J.~L. {Matteson}}, \bibinfo{author}{Y.~{Gu}},
	\bibinfo{author}{R.~T. {Skelton}}, \bibinfo{author}{A.~C. {Deal}},
	\bibinfo{author}{E.~A. {Stephan}}, \bibinfo{author}{F.~{Duttweiler}},
	\bibinfo{author}{G.~L. {Huszar}}, \bibinfo{author}{T.~M. {Gasaway}},
	\bibinfo{author}{C.~S. {Levin}},
	\newblock \bibinfo{title}{Charge collection studies of a high resolution
		czt-based detector for pet},
	\newblock in: \bibinfo{booktitle}{2008 IEEE Nuclear Science Symposium
		Conference Record}, \bibinfo{year}{2008}, pp. \bibinfo{pages}{503--510}.
	\DOIprefix\doi{10.1109/NSSMIC.2008.4775215}.
	\bibitem[{Peng and Levin(2010)}]{Peng2010}
	\bibinfo{author}{H.~Peng}, \bibinfo{author}{C.~S. Levin},
	\newblock \bibinfo{title}{Design study of a high-resolution breast-dedicated
		pet system built from cadmium zinc telluride detectors},
	\newblock \bibinfo{journal}{Physics in medicine and biology}
	\bibinfo{volume}{55} (\bibinfo{year}{2010}) \bibinfo{pages}{2761--2788}.
	\URLprefix \url{https://doi.org/10.1088/0031-9155/55/9/022}.
	\DOIprefix\doi{10.1088/0031-9155/55/9/022}, \bibinfo{note}{20400807[pmid]}.
	\bibitem[{Gu and Levin(2014)}]{Gu_2014}
	\bibinfo{author}{Y.~Gu}, \bibinfo{author}{C.~S. Levin},
	\newblock \bibinfo{title}{Study of electrode pattern design for a {CZT}-based
		{PET} detector},
	\newblock \bibinfo{journal}{Physics in Medicine and Biology}
	\bibinfo{volume}{59} (\bibinfo{year}{2014}) \bibinfo{pages}{2599--2621}.
	\URLprefix \url{https://doi.org/10.1088/0031-9155/59/11/2599}.
	\DOIprefix\doi{10.1088/0031-9155/59/11/2599}.
	\bibitem[{Abbaszadeh and Levin(2017)}]{Abbaszadeh_2017}
	\bibinfo{author}{S.~Abbaszadeh}, \bibinfo{author}{C.~S. Levin},
	\newblock \bibinfo{title}{{New-generation small animal positron emission
			tomography system for molecular imaging}},
	\newblock \bibinfo{journal}{Journal of Medical Imaging} \bibinfo{volume}{4}
	(\bibinfo{year}{2017}) \bibinfo{pages}{1 -- 7}. \URLprefix
	\url{https://doi.org/10.1117/1.JMI.4.1.011008}.
	\DOIprefix\doi{10.1117/1.JMI.4.1.011008}.
	\bibitem[{Bugby et~al.(2019)Bugby, Koch-Mehrin, Veale, Wilson, and
		Lees}]{BUGBY2019}
	\bibinfo{author}{S.~Bugby}, \bibinfo{author}{K.~Koch-Mehrin},
	\bibinfo{author}{M.~Veale}, \bibinfo{author}{M.~Wilson},
	\bibinfo{author}{J.~Lees},
	\newblock \bibinfo{title}{Energy-loss correction in charge sharing events for
		improved performance of pixellated compound semiconductors},
	\newblock \bibinfo{journal}{Nuclear Instruments and Methods in Physics Research
		Section A: Accelerators, Spectrometers, Detectors and Associated Equipment}
	\bibinfo{volume}{940} (\bibinfo{year}{2019}) \bibinfo{pages}{142 -- 151}.
	\URLprefix
	\url{http://www.sciencedirect.com/science/article/pii/S0168900219308484}.
	\DOIprefix\doi{https://doi.org/10.1016/j.nima.2019.06.017}.
	\bibitem[{Abbene et~al.(2018)Abbene, Gerardi, Principato, Bettelli, Seller,
		Veale, Fox, Sawhney, Zambelli, Benassi, and Zappettini}]{Abbene2018}
	\bibinfo{author}{L.~Abbene}, \bibinfo{author}{G.~Gerardi},
	\bibinfo{author}{F.~Principato}, \bibinfo{author}{M.~Bettelli},
	\bibinfo{author}{P.~Seller}, \bibinfo{author}{M.~C. Veale},
	\bibinfo{author}{O.~Fox}, \bibinfo{author}{K.~Sawhney},
	\bibinfo{author}{N.~Zambelli}, \bibinfo{author}{G.~Benassi},
	\bibinfo{author}{A.~Zappettini},
	\newblock \bibinfo{title}{{Dual-polarity pulse processing and analysis for
			charge-loss correction in cadmium–zinc–telluride pixel detectors}},
	\newblock \bibinfo{journal}{Journal of Synchrotron Radiation}
	\bibinfo{volume}{25} (\bibinfo{year}{2018}) \bibinfo{pages}{1078--1092}.
	\URLprefix \url{http://scripts.iucr.org/cgi-bin/paper?S1600577518006422}.
	\DOIprefix\doi{10.1107/S1600577518006422}.
	\bibitem[{Barrett et~al.(1995)Barrett, Eskin, and Barber}]{Barret}
	\bibinfo{author}{H.~H. Barrett}, \bibinfo{author}{J.~D. Eskin},
	\bibinfo{author}{H.~B. Barber},
	\newblock \bibinfo{title}{Charge transport in arrays of semiconductor gamma-ray
		detectors},
	\newblock \bibinfo{journal}{Phys. Rev. Lett.} \bibinfo{volume}{75}
	(\bibinfo{year}{1995}) \bibinfo{pages}{156--159}. \URLprefix
	\url{https://link.aps.org/doi/10.1103/PhysRevLett.75.156}.
	\DOIprefix\doi{10.1103/PhysRevLett.75.156}.
	\bibitem[{{Luke}(1996)}]{Luke_1996}
	\bibinfo{author}{P.~{Luke}},
	\newblock \bibinfo{title}{{Electrode configuration and energy resolution in
			gamma-ray detectors}},
	\newblock \bibinfo{journal}{Nuclear Instruments and Methods in Physics Research
		A} \bibinfo{volume}{380} (\bibinfo{year}{1996}) \bibinfo{pages}{232--237}.
	\DOIprefix\doi{https://doi.org/10.1016/S0168-9002(96)00353-1.}
	\bibitem[{{Kuvvetli} and {Budtz-Jorgensen}(2005)}]{Kuvvetli_2005}
	\bibinfo{author}{I.~{Kuvvetli}}, \bibinfo{author}{C.~{Budtz-Jorgensen}},
	\newblock \bibinfo{title}{Pixelated cdznte drift detectors},
	\newblock \bibinfo{journal}{IEEE Transactions on Nuclear Science}
	\bibinfo{volume}{52} (\bibinfo{year}{2005}) \bibinfo{pages}{1975--1981}.
	\DOIprefix\doi{10.1109/TNS.2005.856882}.
	\bibitem[{Abbene et~al.(2007)Abbene, {Del Sordo}, Fauci, Gerardi, {La Manna},
		Raso, Cola, Perillo, Raulo, Gostilo, and Stumbo}]{Abbene_2007}
	\bibinfo{author}{L.~Abbene}, \bibinfo{author}{S.~{Del Sordo}},
	\bibinfo{author}{F.~Fauci}, \bibinfo{author}{G.~Gerardi},
	\bibinfo{author}{A.~{La Manna}}, \bibinfo{author}{G.~Raso},
	\bibinfo{author}{A.~Cola}, \bibinfo{author}{E.~Perillo},
	\bibinfo{author}{A.~Raulo}, \bibinfo{author}{V.~Gostilo},
	\bibinfo{author}{S.~Stumbo},
	\newblock \bibinfo{title}{Spectroscopic response of a cdznte multiple electrode
		detector},
	\newblock \bibinfo{journal}{Nuclear Instruments and Methods in Physics Research
		Section A: Accelerators, Spectrometers, Detectors and Associated Equipment}
	\bibinfo{volume}{583} (\bibinfo{year}{2007}) \bibinfo{pages}{324--331}.
	\URLprefix
	\url{https://www.sciencedirect.com/science/article/pii/S0168900207019924}.
	\DOIprefix\doi{https://doi.org/10.1016/j.nima.2007.09.015}.
	\bibitem[{Duff et~al.(2008)Duff, Burger, Groza, Buliga, Bradley, Dai, Teslich,
		Awadalla, Mackenzie, and Chen}]{Duff2008}
	\bibinfo{author}{M.~C. Duff}, \bibinfo{author}{A.~Burger},
	\bibinfo{author}{M.~Groza}, \bibinfo{author}{V.~Buliga},
	\bibinfo{author}{J.~P. Bradley}, \bibinfo{author}{Z.~R. Dai},
	\bibinfo{author}{N.~Teslich}, \bibinfo{author}{S.~A. Awadalla},
	\bibinfo{author}{J.~Mackenzie}, \bibinfo{author}{H.~Chen},
	\newblock \bibinfo{title}{{Characterization of detector grade CdZnTe material
			from Redlen Technologies}},
	\newblock \bibinfo{journal}
	\bibinfo{volume}{7079} (\bibinfo{year}{2008}) \bibinfo{pages}{223--237}.
	\DOIprefix\doi{10.1117/12.798921}.
	\bibitem[{{Tumer} et~al.(2008){Tumer}, {Cajipe}, {Clajus}, {Hayakawa}, and
		{Volkovskii}}]{Tumer08}
	\bibinfo{author}{T.~O. {Tumer}}, \bibinfo{author}{V.~B. {Cajipe}},
	\bibinfo{author}{M.~{Clajus}}, \bibinfo{author}{S.~{Hayakawa}},
	\bibinfo{author}{A.~{Volkovskii}},
	\newblock \bibinfo{title}{{Performance of RENA-3 IC with position-sensitive
			solid-state detectors}},
	\newblock in: \bibinfo{booktitle}{Hard X-Ray, Gamma-Ray, and Neutron Detector
		Physics X}, volume \bibinfo{volume}{7079} of \textit{\bibinfo{series}{"Proc.
			SPIE"}}, \bibinfo{year}{2008}, p. \bibinfo{pages}{70791F}.
	\DOIprefix\doi{10.1117/12.797750}.
	\bibitem[{{Li} et~al.(2000){Li}, {He}, {Knoll}, {Wehe}, and
		{Du}}]{DOI_W_Li_1999}
	\bibinfo{author}{W.~{Li}}, \bibinfo{author}{Z.~{He}}, \bibinfo{author}{G.~F.
		{Knoll}}, \bibinfo{author}{D.~K. {Wehe}}, \bibinfo{author}{Y.~F. {Du}},
	\newblock \bibinfo{title}{A modeling method to calibrate the interaction depth
		in 3-d position sensitive cdznte gamma-ray spectrometers},
	\newblock \bibinfo{journal}{IEEE Transactions on Nuclear Science}
	\bibinfo{volume}{47} (\bibinfo{year}{2000}) \bibinfo{pages}{890--894}.
	\DOIprefix\doi{10.1109/23.856713}.
	\bibitem[{{Kalemci} et~al.(1999){Kalemci}, {Matteson}, {Skelton}, {Hink}, and
		{Slavis}}]{Kalemci99}
	\bibinfo{author}{E.~{Kalemci}}, \bibinfo{author}{J.~L. {Matteson}},
	\bibinfo{author}{R.~T. {Skelton}}, \bibinfo{author}{P.~L. {Hink}},
	\bibinfo{author}{K.~R. {Slavis}},
	\newblock \bibinfo{title}{{Model calculations of the response of CZT strip
			detectors}},
	\newblock in: \bibinfo{editor}{R.~B. {James}}, \bibinfo{editor}{R.~C.
		{Schirato}} (Eds.), \bibinfo{booktitle}{Hard X-Ray, Gamma-Ray, and Neutron
		Detector Physics}, volume \bibinfo{volume}{3768} of
	\textit{\bibinfo{series}{"Proc. SPIE"}}, \bibinfo{year}{1999}, pp.
	\bibinfo{pages}{360--373}.
	\bibitem[{He(2001)}]{HE_2001}
	\bibinfo{author}{Z.~He},
	\newblock \bibinfo{title}{Review of the shockley–ramo theorem and its
		application in semiconductor gamma-ray detectors},
	\newblock \bibinfo{journal}{Nuclear Instruments and Methods in Physics Research
		Section A: Accelerators, Spectrometers, Detectors and Associated Equipment}
	\bibinfo{volume}{463} (\bibinfo{year}{2001}) \bibinfo{pages}{250 -- 267}.
	\URLprefix
	\url{http://www.sciencedirect.com/science/article/pii/S0168900201002236}.
	\DOIprefix\doi{https://doi.org/10.1016/S0168-9002(01)00223-6}.
	\bibitem[{S.~Agostinelli(2003)}]{Agostinelli_G4_2003}
	\bibinfo{author}{e.~a. S.~Agostinelli},
	\newblock \bibinfo{title}{Geant4—a simulation toolkit},
	\newblock \bibinfo{journal}{Nuclear Instruments and Methods in Physics Research
		Section A: Accelerators, Spectrometers, Detectors and Associated Equipment}
	\bibinfo{volume}{506} (\bibinfo{year}{2003}) \bibinfo{pages}{250 -- 303}.
	\URLprefix
	\url{http://www.sciencedirect.com/science/article/pii/S0168900203013688}.
	\DOIprefix\doi{https://doi.org/10.1016/S0168-9002(03)01368-8}.
	\bibitem[{{Benoit} and {Hamel}(2009)}]{Benoit09}
	\bibinfo{author}{M.~{Benoit}}, \bibinfo{author}{L.~A. {Hamel}},
	\newblock \bibinfo{title}{{Simulation of charge collection processes in
			semiconductor CdZnTe {$\gamma$}-ray detectors}},
	\newblock \bibinfo{journal}{Nuclear Instruments and Methods in Physics Research
		A} \bibinfo{volume}{606} (\bibinfo{year}{2009}) \bibinfo{pages}{508--516}.
	\DOIprefix\doi{10.1016/j.nima.2009.04.019}.
	\bibitem[{Bolotnikov et~al.(1999)Bolotnikov, Cook, Harrison, Wong, Schindler,
		and Eichelberger}]{Bolotnikov_1999}
	\bibinfo{author}{A.~Bolotnikov}, \bibinfo{author}{W.~Cook},
	\bibinfo{author}{F.~Harrison}, \bibinfo{author}{A.-S. Wong},
	\bibinfo{author}{S.~Schindler}, \bibinfo{author}{A.~Eichelberger},
	\newblock \bibinfo{title}{Charge loss between contacts of cdznte pixel
		detectors},
	\newblock \bibinfo{journal}{Nuclear Instruments and Methods in Physics Research
		Section A: Accelerators, Spectrometers, Detectors and Associated Equipment}
	\bibinfo{volume}{432} (\bibinfo{year}{1999}) \bibinfo{pages}{326 -- 331}.
	\URLprefix
	\url{http://www.sciencedirect.com/science/article/pii/S0168900299003885}.
	\DOIprefix\doi{https://doi.org/10.1016/S0168-9002(99)00388-5}.
	\bibitem[{{Luke} et~al.(2001){Luke}, {Lee}, {Amman}, and {Yu}}]{Luke_2001}
	\bibinfo{author}{P.~N. {Luke}}, \bibinfo{author}{J.~S. {Lee}},
	\bibinfo{author}{M.~{Amman}}, \bibinfo{author}{K.~M. {Yu}},
	\newblock \bibinfo{title}{Noise reduction in cdznte coplanar-grid detectors},
	\newblock in: \bibinfo{booktitle}{2001 IEEE Nuclear Science Symposium
		Conference Record (Cat. No.01CH37310)}, volume~\bibinfo{volume}{4},
	\bibinfo{year}{2001}, pp. \bibinfo{pages}{2272--2275 vol.4}.
	\DOIprefix\doi{10.1109/NSSMIC.2001.1009276}.
	\bibitem[{{Veale} et~al.(2011){Veale}, {Bell}, {Jones}, {Seller}, {Wilson},
		{Allwork}, {Kitou}, {Sellin}, {Veeramani}, and {Cernik}}]{veale_2011}
	\bibinfo{author}{M.~C. {Veale}}, \bibinfo{author}{S.~J. {Bell}},
	\bibinfo{author}{L.~L. {Jones}}, \bibinfo{author}{P.~{Seller}},
	\bibinfo{author}{M.~D. {Wilson}}, \bibinfo{author}{C.~{Allwork}},
	\bibinfo{author}{D.~{Kitou}}, \bibinfo{author}{P.~J. {Sellin}},
	\bibinfo{author}{P.~{Veeramani}}, \bibinfo{author}{R.~C. {Cernik}},
	\newblock \bibinfo{title}{An asic for the study of charge sharing effects in
		small pixel cdznte x-ray detectors},
	\newblock \bibinfo{journal}{IEEE Transactions on Nuclear Science}
	\bibinfo{volume}{58} (\bibinfo{year}{2011}) \bibinfo{pages}{2357--2362}.
	\DOIprefix\doi{10.1109/TNS.2011.2162746}.
	\bibitem[{{Kim} et~al.(2009){Kim}, {Anderson}, {Kaye}, {Kaye}, {Zhu}, {Zhang},
		and {He}}]{Kim_200Conf}
	\bibinfo{author}{J.~C. {Kim}}, \bibinfo{author}{S.~E. {Anderson}},
	\bibinfo{author}{W.~{Kaye}}, \bibinfo{author}{S.~J. {Kaye}},
	\bibinfo{author}{Y.~{Zhu}}, \bibinfo{author}{F.~{Zhang}},
	\bibinfo{author}{Z.~{He}},
	\newblock \bibinfo{title}{Study on effect of charge sharing events in
		common-grid pixelated cdznte detectors},
	\newblock in: \bibinfo{booktitle}{2009 IEEE Nuclear Science Symposium
		Conference Record (NSS/MIC)}, \bibinfo{year}{2009}, pp.
	\bibinfo{pages}{1640--1646}. \DOIprefix\doi{10.1109/NSSMIC.2009.5402243}.
	\bibitem[{Bolotnikov et~al.(2002)Bolotnikov, Chen, Cook, Harrison, Kuvvetli,
		and Schindler}]{Bolotnikov2002}
	\bibinfo{author}{A.~E. Bolotnikov}, \bibinfo{author}{C.~M. Chen},
	\bibinfo{author}{W.~R. Cook}, \bibinfo{author}{F.~A. Harrison},
	\bibinfo{author}{I.~Kuvvetli}, \bibinfo{author}{S.~M. Schindler},
	\newblock \bibinfo{title}{{Effects of bulk and surface conductivity on the
			performance of CdZnTe pixel detectors}},
	\newblock \bibinfo{journal}{IEEE Transactions on Nuclear Science}
	\bibinfo{volume}{49 I} (\bibinfo{year}{2002}) \bibinfo{pages}{1941--1949}.
	\DOIprefix\doi{10.1109/TNS.2002.801673}.
	\bibitem[{Boiotnikov et~al.(2005)Boiotnikov, Camarda, Wright, and
		James}]{Bolotnikov2005}
	\bibinfo{author}{A.~E. Boiotnikov}, \bibinfo{author}{G.~C. Camarda},
	\bibinfo{author}{G.~W. Wright}, \bibinfo{author}{R.~B. James},
	\newblock \bibinfo{title}{{Factors limiting the performance of CdZnTe
			detectors}},
	\newblock \bibinfo{journal}{IEEE Transactions on Nuclear Science}
	\bibinfo{volume}{52} (\bibinfo{year}{2005}) \bibinfo{pages}{589--598}.
	\DOIprefix\doi{10.1109/TNS.2005.851419}.
	\bibitem[{Kamieniecki(2014)}]{Kamieniecki2014}
	\bibinfo{author}{E.~Kamieniecki},
	\newblock \bibinfo{title}{{Effect of charge trapping on effective carrier
			lifetime in compound semiconductors: High resistivity CdZnTe}},
	\newblock \bibinfo{journal}{Journal of Applied Physics} \bibinfo{volume}{116}
	(\bibinfo{year}{2014}) \bibinfo{pages}{193702}. \URLprefix
	\url{https://aip.scitation.org/doi/abs/10.1063/1.4901826}.
	\DOIprefix\doi{10.1063/1.4901826}.
	
\end{thebibliography}
\end{document}